\documentclass[prd,amsmath,floats,amssymb,authoryear,floatfix,superscriptaddress,nofootinbib,onecolumn]{revtex4} 

\usepackage{graphicx}
\usepackage{hhline} 
\usepackage{placeins}
\usepackage{caption}
\usepackage{comment}
\usepackage{ulem}
\usepackage{natbib}
\usepackage{multirow}
\usepackage{makecell}
\usepackage{placeins}

\renewcommand{\thetable}{\textbf{\arabic{table}}}

\makeatletter
\renewcommand{\fnum@figure}{\textbf{Figure~\thefigure}}
\renewcommand{\fnum@table}{\textbf{Table~\thetable}}
\makeatother

\newcommand{\rthis}[1]{\textcolor{black}{#1}}
\usepackage{float}
\usepackage[caption=false]{subfig}

\usepackage[plainpages=false, colorlinks=true, anchorcolor=blue, linkcolor=blue, citecolor=blue, bookmarks=false]{hyperref}

\begin{document}
\preprint{APS/123-QED}

\title{Impact of CMB low-$\ell$ EE polarization data on dark energy parameterizations}

\author{Shubham Barua}
 \altaffiliation{Email:ph24resch01006@iith.ac.in}
\author{Shantanu Desai}
 \altaffiliation{Email:shntn05@gmail.com}
\affiliation{
 Department of Physics, IIT Hyderabad Kandi, Telangana 502284,  India}

\begin{abstract}
Measurement of the optical depth to reionization ($\tau_\mathrm{reio}$) is largely driven by the large-scale EE polarization data of CMB ($\ell<30$). Removing the low-$\ell$ EE data potentially alleviates various cosmological tensions. In this work, we study the effect of the low-$\ell$ EE polarization measurements on the CPL, JBP and BA dark energy parameterizations using CMB data from Planck and ACT DR6, combined with DESI BAO and PantheonPlus compilation of Type Ia supernovae. We find that excluding low-$\ell$ EE data shifts $\tau_\mathrm{reio}$ and $A_s$ to higher values through the unbroken $A_s-\tau_\mathrm{reio}$ degeneracy with a $\sim(1.4-1.8)\sigma$ shift in $A_s$ for $\Lambda$CDM and JBP,  and a milder $\sim 1\sigma$ shift for CPL and BA. The equation of state (EOS) for all three parameterizations moves towards the quintessence regime ($w(z) > -1$) upon exclusion of low-$\ell$ EE data, driven primarily by the strengthening of the $w_a-A_s$ and the $w_a-\tau_\mathrm{reio}$ correlations, with a small effect from the correlations of $w_0$ with $A_s$ and $\tau_\mathrm{reio}$. The most prominent effect occurs in JBP, where the EOS lies entirely within the quintessence regime at $1\sigma$ when excluding low-$\ell$ EE data. Model comparison through AIC shows positive evidence in favor of CPL and BA and weak evidence in favor of JBP, robust to the inclusion of low-$\ell$ EE data and CMB data used. DIC model comparison shows \rthis{strong evidence in favor of CPL and BA when low-$\ell$ EE data is included, and positive evidence when it is excluded}. For JBP, we get positive (weak) evidence when including (excluding) low-$\ell$ EE data \rthis{over $\Lambda$CDM}. 
\end{abstract}


\maketitle
\section{Introduction}
\label{sec:1}

Reionization marks the epoch in the cosmic history when radiation from the first celestial objects in the universe ionized the neutral intergalactic medium~\cite{robertson_2010, loeb_2001, choudhury_2022}. Estimating the timeline of this event remains one of the outstanding challenges of cosmology, as  it provides crucial insights into the  onset of  star and galaxy formation, through the reionization of baryonic matter.
The history of reionization has been studied extensively using various observations. These include quasar and galaxy spectra~\cite{mesinger_2007, davies_2018, mason_2018, greig_2022, umeda_2024, elbers_2025}, the fraction of dark pixels in Lyman-$\alpha$ and Lyman-$\beta$ forests~\cite{jin_2023, mcgreer_2015}, clustering of Lyman-$\alpha$ emitters~\cite{sobacchi_2015, ouchi_2010}, the kinematic Sunyaev-Zeldovich (kSZ) effect~\cite{zahn_2012, smith_2017, raghunathan_2024}, post-reionization Lyman-$\alpha$ forest~\cite{cen_2009} and the 21-cm power spectrum~\cite{zara_2022, ghara_2025}. 

Telescopes such as the James Webb Space Telescope (JWST)~\cite{jakobsen_2022, wright_2023}, Atacama Cosmology Telescope (ACT)~\cite{act_2025, maccrann_2024, qu_2024, madhavacheril_2024} and the South Pole Telescope (SPT)~\cite{spt_2026, carlstrom_2011} and space-based satellites such as Planck~\cite{planck_2020} have enabled us to explore this epoch in more detail~\cite{qu_2026}. JWST data suggests a high value of the optical depth to reionization ($\tau_\mathrm{reio} \gtrsim 0.07$)~\cite{munoz_2024, curtis_2023, llerena_2025, umeda_2026, cohon_2025}. Planck, on the other hand, has measured $\tau_\mathrm{reio} = 0.0544 \pm 0.0073$~\cite{planck_2020}. 

During the reionization epoch, CMB photons undergo Thomson scattering off free electrons. Consequently, their trajectories \rthis{are altered}. This leaves an imprint in the angular power spectra of temperature and polarization anisotropies. This effect is observed as a suppression of the CMB spectra by a factor of $e^{-\tau_\mathrm{reio}}$ at multipoles $\ell \gtrsim 100$. This is degenerate with a shift in the primordial power spectrum amplitude $A_s$. During reionization, the large-scale modes outside the horizon experience less suppression. Additionally, Thomson scattering in the late universe enhances the E-mode spectra for $\ell \lesssim 10$. These effects break the degeneracy between $A_s$ and $\tau_\mathrm{reio}$. 

The precise determination of $\tau_\mathrm{reio}$ from CMB relies largely on the large-scale polarization data, which contain the largest uncertainties among CMB data \cite{giare_2024}. When removing the large-scale CMB data, $\tau_{\mathrm{reio}}$ is consistent with the values from JWST data~\cite{endsley_2024, simmonds_2024}. Several works have tried to infer constraints on $\tau_\mathrm{reio}$ using several independent methods to minimize the dependence on a single dataset~\cite{mcquinn_2016, robertson_2015, elbers_2025}.

Recent baryon acoustic oscillation (BAO) measurements from the Dark Energy Spectroscopic Instrument (DESI) data release 2~\cite{desi_2025} combined with cosmic microwave background (CMB) data and/or Type Ia Supernova (SNe Ia) data have  provided evidence for dynamical dark energy (DDE)~\cite{garcia_2025, yang_2019, li_2024, choudhury_2025, choudhury_2024, choudhury_2025a, park_2024, herold_2025, park_2025, perez_2024, Li_2026} when considering the Chevallier-Polarski-Linder (CPL) parameterization~\cite{chevallier_2001, linder_2003}. This has been noted to be in part due to the tension between the low-intermediate $z$ BAO data and high $z$ CMB data. This tension, as pointed out in~\cite{sailer_2025, jhaveri_2025, upadhyay_2026} could be explained by unknown systematics in large-scale polarization measurements of $\tau_\mathrm{reio}$.

To examine whether and how much the $\tau_\mathrm{reio}$ sensitivity persists across different DE parameterizations and motivated by \cite{allali_2025} and \cite{sailer_2025}, in this work we extend the analysis to include the Jassal-Bagla-Padmanabhan (JBP)~\cite{jbp_2005} and Barboza-Alcaniz (BA)~\cite{ba_2008} DE parameterizations along with the standard $\Lambda$CDM model and the CPL parameterizations. It has been seen in \cite{sailer_2025} that using Planck high-$\ell$ data along with different Gaussian priors on $\tau_\mathrm{reio}$ (instead of using low-$\ell$ data), the evidence for evolving DE reduces to a mild 1.4$\sigma$. This occurs when using $\tau_\mathrm{reio} = 0.09$.

In this work, we study the parameterizations by constraining them with both Planck and ACT DR6 CMB data. Usually, when working with ACT primary CMB likelihood, we utilize either a Gaussian prior on $\tau_\mathrm{reio}$ or combine it with low-$\ell$ Planck likelihood, since ACT itself does not measure low-$\ell$ EE polarization data. In our case, when considering the ACT CMB data, we either use low-$\ell$ Planck likelihood or use only ACT data and let the $\tau_\mathrm{reio}$ be constrained through parameter degeneracies, primarily with $A_s$.

The paper is organized as follows. In Section~\ref{sec:2} and~\ref{sec:3}, we describe the DE parameterizations and the methodology used in this work. In Section~\ref{sec:4}, we present our results and associated discussions. We conclude in Section~\ref{sec:5}.

\section{Dark Energy Parameterizations}
\label{sec:2}

The expansion history of the universe can be written as~\cite{zheng_2025}:
\begin{equation}
\label{eqn1}
\begin{split}
    H^2(z) = H_0^2\Big[\Omega_r(1+z)^4 + \Omega_m(1+z)^3 + f_{DE}(z) + \Omega_k(1+z)^2\Big],
\end{split}
\end{equation}
where $\Omega_r$, $\Omega_m$, and $\Omega_k$ represent the present dimensionless radiation, matter and curvature densities of the universe, respectively;  $f_{DE}$ is the term representing the dark energy (DE) dynamics given by:
\begin{equation}
\label{eqn2}
    f_{DE}(z) = \Omega_{DE}\text{exp}{\left(3 \int_0^z \frac{1+w(z')}{1+z'}dz'\right)},
\end{equation}
where $\Omega_{DE} = 1 - (\Omega_r+\Omega_m+\Omega_k)$. $\Omega_m$ consists of the baryonic matter density ($\Omega_b$) and the cold dark matter density ($\Omega_c$). In this work, we consider a spatially-flat universe, so $\Omega_k = 0$ henceforth. We utilize three parameterizations of DE defined by the equation of state (EOS) $w(z)$.

\begin{itemize}
    \item \textbf{$\Lambda$CDM}: The DE EOS for $\Lambda$CDM is a constant given by $w(z) = -1$.
    \item \textbf{CPL}: The DE EOS is given by $w(z)=w_0+w_a\frac{z}{1+z}$~\cite{chevallier_2001, linder_2003}. We also apply the condition $w_0 + w_a < 0$ so that DE is subdominant at early times. \rthis{For $z \rightarrow \infty$, $w(z) \rightarrow w_0 + w_a$.}
    \item \textbf{JBP}: DE EOS for this parameterization is $w(z)=w_0+w_a\frac{z}{(1+z)^2}$~\cite{jbp_2005}. \rthis{For $z \rightarrow \infty$, $w(z) \rightarrow w_0$. Therefore, this parameterization peaks at intermediate redshifts while having $w(z) = w_0$ at $z=0$ and $z \rightarrow \infty$.}
    \item \textbf{BA}: The DE EOS is $w(z) = w_0+w_a\frac{z(1+z)}{1+z^2}$~\cite{ba_2008}. \rthis{This parameterization behaves similarly to CPL when $z \rightarrow \infty$ but has a different functional form for $w(z)$,  which is pronounced at intermediate redshifts leading to different cosmological constraints.}
\end{itemize}
\rthis{We consider these three DE parameterizations in order to examine whether the sensitivity of the DE EOS to low-$\ell$ EE polarization data and the shift towards the quintessence regime upon its exclusion (cf.~\cite{giare_2024}) is a generic feature of two-parameter DE EOS or depends on the specific redshift dependence. The differences in the functional form of the three DE parameterizations allow us to test whether the observed $A_s-\tau_\mathrm{reio}$ driven shift in the EOS is a robust feature of DE parameterizations in general or whether it is specific to CPL-like EOS. A comparison of the impact of low-$\ell$ EE polarization data across these functionally different DE parameterizations constitutes one of the main objectives of this work.}

\section{Data Analysis Methods}
\label{sec:3}
We describe the datasets used for our analyses.
\begin{itemize}
    \item \textbf{BAO}: We utilize DESI DR2 BAO measurements~\cite{desi_2025} from multiple tracers in seven different redshift bins. These consist of the spherically averaged distance ($D_V$), the Hubble distance ($D_H$), and the comoving angular diameter distance ($D_M$) normalized to $r_d$ and the correlations between them.
    \item \textbf{Type Ia SNe}: We use the uncalibrated PantheonPlus (PP)~\cite{scolnic_2022}. We consider SNe data corresponding to $z \geq 0.01$ to remove strong peculiar velocity dependencies~\cite{brout_2022}.
    \item \textbf{CMB}: For CMB, we consider data from Planck~\cite{planck_2020, akrami_2018} satellite and the Atacama Cosmology Telescope Data Release 6 (ACT DR6)~\cite{act_2025, calabrese_2025}. For Planck, we utilize the public release 3 (PR3)~\cite{aghanim_2020} TT likelihood ($\ell < 30$), the \texttt{CamSpec} likelihood from 2020 public release 4 (PR4)~\cite{efstathiou_2019, rosenberg_2022} for $\ell > 30$ TT, EE and TE data. For the $\ell < 30$ EE polarization data, we employ the 2019 \texttt{sroll2} likelihood~\cite{delouis_2019, pagano_2020}. 
    
    ACT DR6 provides greater sensitivity \rthis{at high multipoles} ($\ell> 1000$), unlike Planck, which covers low to intermediate $\ell$ better. We use ACT DR6 TT, TE and EE likelihoods at $\ell > 600$. For low-$\ell$ EE, we add the \texttt{sroll2} likelihood~\cite{delouis_2019}.

    For CMB lensing, we use ACT DR6 lensing likelihood~\cite{qu_2024, madhavacheril_2024} combined with the Planck PR4 lensing likelihood~\cite{carron_2022}. This CMB lensing data require an increase in precision settings recommended by the ACT collaboration~\cite{calabrese_2025, herold_2025}. Henceforth, the CMB data combined with CMB lensing shall be referred to as CMB only (CMB lensing is always included in our dataset combinations). We consider \rthis{a neutrino sector consisting of one} massive and two massless neutrinos with $\Sigma m_{\nu} = 0.06$ eV.
\end{itemize}

To study the impact of low-$\ell$ data on DDE parameterizations, we consider two data combinations:
\begin{itemize}
    \item The first one consists of CMB$+$SNe$+$BAO datasets where CMB data is from either Planck or ACT DR6 (including low-$\ell$ EE).
    \item The second one consists of CMB$+$SNe$+$BAO datasets but we remove the low-$\ell$ EE likelihood (provided by \texttt{sroll2}) from the CMB data.
\end{itemize}

Our analysis is carried out using the Boltzmann code \texttt{CLASS}~\cite{class_2011a, class_2011b} (for the linear theory predictions) while \texttt{hmcode}~\cite{hmcode_2015} is utilized for non-linear corrections to the matter power spectrum. For the Bayesian analysis, we employ \texttt{Cobaya}~\cite{cobaya_2021, 2019_ascl}. \texttt{GetDist}~\cite{getdist_2025} is used for analysis and visualization of the Markov Chain Monte Carlo (MCMC) chains.

Model performance is assessed by examining how well each model fits different dataset combinations using the Akaike Information Criterion (AIC) and Deviance Information Criterion (DIC)~\cite{perez_2024, kunz_2006, marek_2007, liddle_2007,Krishak20}. AIC is defined as: 
\begin{equation}
\label{eqn3}
    \mathrm{AIC} = \chi^2_\mathrm{min} + 2n,
\end{equation}
where $\chi^2_\mathrm{min}$ is the minimum value of $\chi^2$ for the best-fit cosmological parameters and $n$ is the number of free model parameters. DIC is a Bayesian model selection criterion defined as
\begin{equation}
\label{eqn4}
    \mathrm{DIC} = \chi^2(\hat{\theta}) + 2p_D,
\end{equation}
where $p_D = \overline{\chi^2} - \chi^2(\hat{\theta})$. Here, $\overline{\chi^2}$ is the average of the $\chi^2$ values estimated from the MCMC chains and $\chi^2(\hat{\theta})$ is the $\chi^2$ value at the best-fit cosmological parameters $\hat{\theta}$. 

When comparing models, we consider $\Lambda$CDM as our reference model and $\Delta$AIC and $\Delta$DIC are defined as:
\begin{align}
    \Delta\mathrm{AIC} &= \chi^2_\mathrm{min}(\mathcal{M}) - \chi^2_\mathrm{min}(\Lambda \mathrm{CDM})+ 2k \label{eqn5} \\
    \Delta\mathrm{DIC} &= \mathrm{DIC}(\mathcal{M}) - \mathrm{DIC}(\Lambda\mathrm{CDM}), \label{eqn6}
\end{align}
where $k$ is the number of additional parameters \footnote{In our case, it is equal to two in all DE parameterizations considered, since the datasets remain same and only two parameters - $w_0$ and $w_a$ - are added.} in $\mathcal{M}$ compared to $\Lambda$CDM and $\mathcal{M}$ is the model under consideration which is CPL, JBP or BA (in our case). Following standard interpretation~\cite{liddle_2007}, when $-2\leq\Delta\mathrm{AIC},\Delta\mathrm{DIC}<0$, the evidence is \textit{weak} in favor of the model, for $-6\leq\Delta\mathrm{AIC},\Delta\mathrm{DIC}<-2$, the evidence is \textit{positive}, for $-10\leq\Delta\mathrm{AIC},\Delta\mathrm{DIC}<-6$, there is \textit{strong} evidence while for $\Delta\mathrm{AIC},\Delta\mathrm{DIC}<-10$, the evidence in favor of the model under consideration is \textit{very strong}. The sign of $\Delta\mathrm{AIC}$ and $\Delta\mathrm{DIC}$ determines if $\Lambda$CDM is preferred or the model under study is preferred.

\begin{table}[htbp!]
\caption{Priors for cosmological parameters in Eqn.~\ref{eqn1}.}
\label{table1}
\centering
    \begin{tabular}{c|c}
    \hline
    \thead{Parameters} & \thead{Priors} \\
    \hline
    \hline
    \rule{0pt}{3ex}$100\Omega_bh^2$ & $\mathcal{U}[0.017, 0.027]$ \\[0.5ex]
    $100\Omega_ch^2$ & $\mathcal{U}[0.09, 0.15]$ \\[0.5ex]
    $100\theta_s$ & $\mathcal{U}[0.8, 1.2]$ \\[0.5ex]
    $\ln(10^{10}A_s)$ & $\mathcal{U}[2.6, 3.5]$ \\[0.5ex]
    $n_s$ & $\mathcal{U}[0.9, 1.1]$ \\[0.5ex]
    $\tau_\mathrm{reio}$ & $\mathcal{U}[0.03, 0.1]$ \\[0.5ex] 
    $w_0$ & $\mathcal{U}[-3, 1]$ \\[0.5ex]
    $w_a$ & $\mathcal{U}[-3, 2]$ \\[0.5ex]
    \hline
    \end{tabular}
\end{table}

\section{Results and discussion}
\label{sec:4}

The $68\%$ credible intervals for the cosmological parameters obtained using Bayesian analysis are summarized in Tables~\ref{table2}-\ref{table3}. The $\Delta$AIC and $\Delta$DIC used for model comparison are presented in Fig.~\ref{fig7}. In Fig.~\ref{fig8}, we present the likelihood contributions from the constituent datasets to the total likelihood for ACT DR6 and Planck CMB data, respectively. The $\chi^2$ values are listed in Tables~\ref{table4}-\ref{table5} in Appendix~\ref{appB}. In Figs.~\ref{fig2}-\ref{fig4}, we present the $68\%$ and $95\%$ credible intervals for different dataset-parameterization combinations as well as the EOS plots. In order to understand parameter degeneracies, we display the correlation heatmaps for the CMB$+$BAO$+$SNe dataset (Fig.~\ref{fig5}).

\begin{table}[htbp!]
\caption{$68\%$ credible intervals for cosmological parameters for ACT DR6 CMB$+$BAO$+$SNe dataset combination.}
\label{table2}
\centering
\resizebox{\textwidth}{!}{
    \begin{tabular}{l|c|c|c|c|c|c|c|c}
    \hline
    \thead{Parameters} & \multicolumn{2}{c|}{\thead{$\Lambda$CDM}} & \multicolumn{2}{c|}{\thead{CPL}} & \multicolumn{2}{c|}{\thead{JBP}} & \multicolumn{2}{c}{\thead{BA}} \\
    & \thead{low-$l$ EE} & \thead{no low-$l$ EE} & \thead{low-$l$ EE} & \thead{no low-$l$ EE} & \thead{low-$l$ EE} & \thead{no low-$l$ EE} & \thead{low-$l$ EE} & \thead{no low-$l$ EE} \\
    \hline
    \hline
    \rule{0pt}{3ex}$100\Omega_bh^2$ & $2.263\pm0.016$ & $2.263\pm0.016$ & $2.259\pm0.016$ & $2.260\pm0.016$ & $2.260\pm0.015$ & $2.263\pm0.016$ & $2.261^{+0.016}_{-0.015}$ & $2.260\pm0.016$ \\[1.5ex]
    $100\Omega_ch^2$ & $11.782\pm0.069$ & $11.735\pm0.073$ & $11.93\pm0.10$ & $11.83\pm0.13$ & $11.861\pm0.098$ & $11.74\pm0.11$ & $11.92\pm0.10$ & $11.84\pm0.13$ \\[1.5ex]
    $n_s$ & $0.9758\pm0.0061$ & $0.9784\pm0.0064$ & $0.9739\pm0.0063$ & $0.9773\pm0.0066$ & $0.9749\pm0.0063$ & $0.9785\pm0.0063$ & $0.9738\pm0.0064$ & $0.9772^{+0.0071}_{-0.0064}$ \\[1.5ex]
    $A_s (10^{-9})$ & $2.136^{+0.023}_{-0.025}$ & $2.212^{+0.044}_{-0.027}$ & $2.114^{+0.022}_{-0.025}$ & $2.177^{+0.062}_{-0.043}$ & $2.124^{+0.022}_{-0.027}$ & $2.207^{+0.048}_{-0.030}$ & $2.117^{+0.023}_{-0.026}$ & $2.175^{+0.061}_{-0.044}$ \\[1.5ex]
    $\tau_\mathrm{reio}$ & $0.0644^{+0.0057}_{-0.0069}$ & $>0.0821$ & $0.0599\pm0.0061$ & $0.077^{+0.020}_{-0.009}$ & $0.0617^{+0.0055}_{-0.0069}$ & $>0.0806$ & $0.0602^{+0.0058}_{-0.0066}$ & $0.0767^{+0.019}_{-0.0097}$ \\[1.5ex]
    $z_\mathrm{reio}$ & $8.57^{+0.56}_{-0.65}$ & $10.5^{+1.1}_{-0.47}$ & $8.15\pm0.60$ & $9.71^{+1.60}_{-0.80}$ & $8.32^{+0.55}_{-0.65}$ & $10.4^{+1.2}_{-0.50}$ & $8.18\pm0.61$ & $9.68^{+1.6}_{-0.80}$ \\[1.5ex]
    $H_0$[km/s/Mpc] & $68.38\pm0.28$ & $68.55\pm0.28$ & $67.72\pm0.60$ & $67.65^{+0.57}_{-0.64}$ & $67.70\pm0.58$ & $67.57\pm0.59$ & $67.73^{+0.53}_{-0.60}$ & $67.67\pm0.59$ \\[1.5ex]
    $\sigma_8$ & $0.8130\pm0.0048$ & $0.8264^{+0.0081}_{-0.0051}$ & $0.8163\pm0.0088$ & $0.8206\pm0.0094$ & $0.8118\pm0.0087$ & $0.8178\pm0.0092$ & $0.8159^{+0.0099}_{0.0087}$ & $0.8210\pm0.0098$ \\[1.5ex]
    $w_0$ & $-$ & $-$ & $-0.836\pm0.055$ & $-0.858\pm0.056$ & $-0.807^{+0.073}_{-0.081}$ & $-0.841\pm0.077$ & $-0.858\pm0.049$ & $-0.873\pm0.049$ \\[1.5ex]
    $w_a$ & $-$ & $-$ & $-0.62^{+0.22}_{-0.20}$ & $-0.47^{+0.24}_{-0.21}$ & $-1.13^{+0.50}_{-0.44}$ & $-0.80\pm0.47$ & $-0.31\pm0.11$ & $-0.24^{+0.12}_{-0.10}$ \\[1.5ex]
    \hline
    \end{tabular}
    }
\end{table}

\begin{table}[htbp!]
\caption{$68\%$ credible intervals for cosmological parameters for Planck CMB$+$BAO$+$SNe dataset combination.}
\label{table3}
\centering
\resizebox{\textwidth}{!}{
    \begin{tabular}{l|c|c|c|c|c|c|c|c}
    \hline
    \thead{Parameters} & \multicolumn{2}{c|}{\thead{$\Lambda$CDM}} & \multicolumn{2}{c|}{\thead{CPL}} & \multicolumn{2}{c|}{\thead{JBP}} & \multicolumn{2}{c}{\thead{BA}} \\
    & \thead{low-$l$ EE} & \thead{no low-$l$ EE} & \thead{low-$l$ EE} & \thead{no low-$l$ EE} & \thead{low-$l$ EE} & \thead{no low-$l$ EE} & \thead{low-$l$ EE} & \thead{no low-$l$ EE} \\
    \hline
    \hline
    \rule{0pt}{3ex}$100\Omega_bh^2$ & $2.233\pm0.011$ & $2.237\pm0.012$ & $2.226\pm0.013$ & $2.232\pm0.013$ & $2.229\pm0.013$ & $2.235\pm0.013$ & $2.226\pm0.012$ & $2.231\pm0.013$ \\[1.5ex]  
    $100\Omega_ch^2$ & $11.781\pm0.062$ & $11.752^{+0.059}_{-0.073}$ & $11.89\pm0.081$ & $11.812\pm0.092$ & $11.845\pm0.079$ & $11.769\pm0.087$ & $11.888\pm0.083$ & $11.822\pm0.092$ \\[1.5ex]
    $n_s$ & $0.9684\pm0.0033$ & $0.9696\pm0.0034$ & $0.9657\pm0.0036$ & $0.9681\pm0.0039$ & $0.9669\pm0.0036$ & $0.9695\pm0.0038$ & $0.9657\pm0.0037$ & $0.9679\pm0.0039$ \\[1.5ex]
    $A_s (10^{-9})$ & $2.127^{+0.023}_{-0.026}$ & $2.195^{+0.057}_{-0.025}$ & $2.112^{+0.021}_{-0.024}$ & $2.170^{+0.057}_{-0.050}$ & $2.120^{+0.021}_{-0.026}$ & $2.193^{+0.057}_{-0.035}$ & $2.111^{+0.022}_{-0.026}$ & $2.166\pm0.046$ \\[1.5ex]
    $\tau_\mathrm{reio}$ & $0.0626^{+0.0057}_{-0.0067}$ & $0.0807^{+0.016}_{-0.0049}$ & $0.0591^{+0.0052}_{-0.0061}$ & $0.075^{+0.017}_{-0.012}$ & $0.0609^{+0.0054}_{-0.0064}$ & $0.0803^{+0.016}_{-0.008}$ & $0.0591^{+0.0055}_{-0.0066}$ & $0.0704\pm0.012$ \\[1.5ex]
    $z_\mathrm{reio}$ & $8.46^{+0.57}_{-0.64}$ & $10.1^{+1.3}_{-0.48}$ & $8.14\pm0.56$ & $9.6^{+1.4}_{-1.1}$ & $8.31\pm0.58$ & $10.1^{+1.4}_{-0.64}$ & $8.14^{+0.56}_{-0.63}$ & $9.5^{+1.2}_{-1.0}$ \\[1.5ex]
    $H_0$[km/s/Mpc] & $68.11\pm0.27$ & $68.27^{+0.32}_{-0.48}$ & $67.53\pm0.61$ & $67.56\pm0.60$ & $67.56\pm0.60$ & $67.47\pm0.60$ & $67.57\pm0.58$ & $67.53\pm0.58$ \\[1.5ex]
    $\sigma_8$ & $0.8100\pm0.0049$ & $0.8220^{+0.0099}_{-0.0047}$ & $0.8125\pm0.0082$ & $0.8177\pm0.0093$ & $0.8100\pm0.0082$ & $0.8171\pm0.0092$ & $0.8126\pm0.0081$ & $0.8176\pm0.0087$ \\[1.5ex]
    $w_0$ & $-$ & $-$ & $-0.842\pm0.054$ & $-0.873^{+0.052}_{-0.059}$ & $-0.808\pm0.078$ & $-0.828\pm0.076$ & $-0.870\pm0.046$ & $-0.881\pm0.048$ \\[1.5ex]
    $w_a$ & $-$ & $-$ & $-0.59^{+0.21}_{-0.19}$ & $-0.43^{+0.23}_{-0.18}$ & $-1.14\pm0.46$ & $-0.93\pm0.46$ & $-0.281^{+0.10}_{-0.095}$ & $-0.23\pm0.10$ \\[1.5ex]
    \hline
    \end{tabular}
    }
\end{table}

From Tables~\ref{table2} and~\ref{table3}, we notice that  $H_0$ remains consistent within $1\sigma$ across all dataset-parameterization combinations, showing no significant sensitivity to the choice of CMB dataset or inclusion/exclusion of low-$\ell$ EE data~\cite{giare_2024}. We note that $n_s$ is insensitive to the inclusion of low-$\ell$ EE data and the DE parameterization considered (within $0.5\sigma$) for a specific dataset combination. A mild downward shift of $\sim(1-1.2)\sigma$ (e.g., $0.9784 \pm 0.0064$ for ACT DR6 and $0.9696 \pm 0.0034$ for Planck considering $\Lambda$CDM without low-$\ell$ EE data) is observed when changing the CMB dataset from ACT DR6 to Planck. The error bars are larger ($\sim 43\%$) in ACT DR6 dataset combinations  compared to Planck~\cite{act_2025}. $\Omega_bh^2$ is also insensitive to the inclusion of low-$\ell$ EE data and the DE parameterization considered. A mild $\sim(1.3-1.8)\sigma$ downward shift (e.g., $(2.260\pm0.015)\times10^{-2}$ for ACT DR6 and $(2.232 \pm 0.013)\times10^{-2}$ for Planck considering JBP with low-$\ell$ EE data) is observed when changing CMB data from ACT DR6 to Planck~\cite{act_2025}. $\Omega_ch^2$ is insensitive to CMB data and inclusion of low-$\ell$ EE data. A mild shift of $\sim(1-1.2) \sigma$ is observed when comparing $\Lambda$CDM with CPL and BA parameterizations for a particular dataset combination. For JBP parameterization, this shift is within $1\sigma$ in all cases. low-$\ell$ EE data has an impact on the error bars of $\Omega_ch^2$. \rthis{Compared to $\Lambda$CDM, the error bars on $\Omega_ch^2$ for CPL, BA, and JBP parameterizations are relaxed by $\sim(40-45)\%$ (with EE) and $\sim(50-80)\%$ (without EE) for ACT DR6. For Planck, the corresponding relaxation is $\sim(25-35)\%$ (with EE) and $\sim(30-40)\%$ (without EE). This indicates that low-$\ell$ EE data plays a non-trivial role in constraining $\Omega_ch^2$ in extended DE models.}

For $A_s$, excluding low-$\ell$ EE data produces an upward shift of $\sim (1.4-1.8)\sigma$ in $\Lambda$CDM and JBP, while for CPL and BA the shift is milder at $\sim 1\sigma$. This \rthis{behaviour} is robust across both CMB datasets, with ACT DR6 (e.g., $(2.136^{+0.023}_{-0.025})\times10^{-9}$ when including low-$\ell$ EE and $(2.212^{+0.044}_{-0.027})\times10^{-9}$ when excluding low-$\ell$ EE data for $\Lambda$CDM) displaying a slightly larger shift than Planck (e.g., $(2.127^{+0.023}_{-0.026})\times10^{-9}$ with low-$\ell$ EE data and $(2.195^{+0.057}_{-0.025})\times10^{-9}$ without low-$\ell$ EE data for $\Lambda$CDM). In all other cases (changing CMB dataset or DE parameterization for the same dataset), $A_s$ remains consistent within $0.5\sigma$. \rthis{We note that ACT DR6's calibration is refined using a comparison with Planck's temperature maps~\cite{act_2025}, so the agreement in $A_s$ between the two CMB datasets may reflect their shared calibration, rather than fully independent constraints.} The $A_s$ error bars relax by $\sim (50-90)\%$ for $\Lambda$CDM and JBP and by $\sim(90-120)\%$ for CPL and BA, when excluding low-$\ell$ EE data. The larger relaxation is consistent with the smaller $A_s$ discrepancies observed in CPL and BA compared to $\Lambda$CDM and JBP. The choice of CMB dataset (Planck or ACT DR6) shows broadly similar relaxation, with Planck showing slightly larger values. There exists a negative correlation between $A_s$ and $\Omega_ch^2$ (cf. Fig.~\ref{fig5}). The strength of this correlation is weak in $\Lambda$CDM but stronger in CPL, JBP and BA. \rthis{It strengthens when excluding large-scale EE polarization data}.

\begin{figure}[H]
    \centering
    \subfloat[ACT DR6\label{1a}]{
    \includegraphics[width=0.5\textwidth]{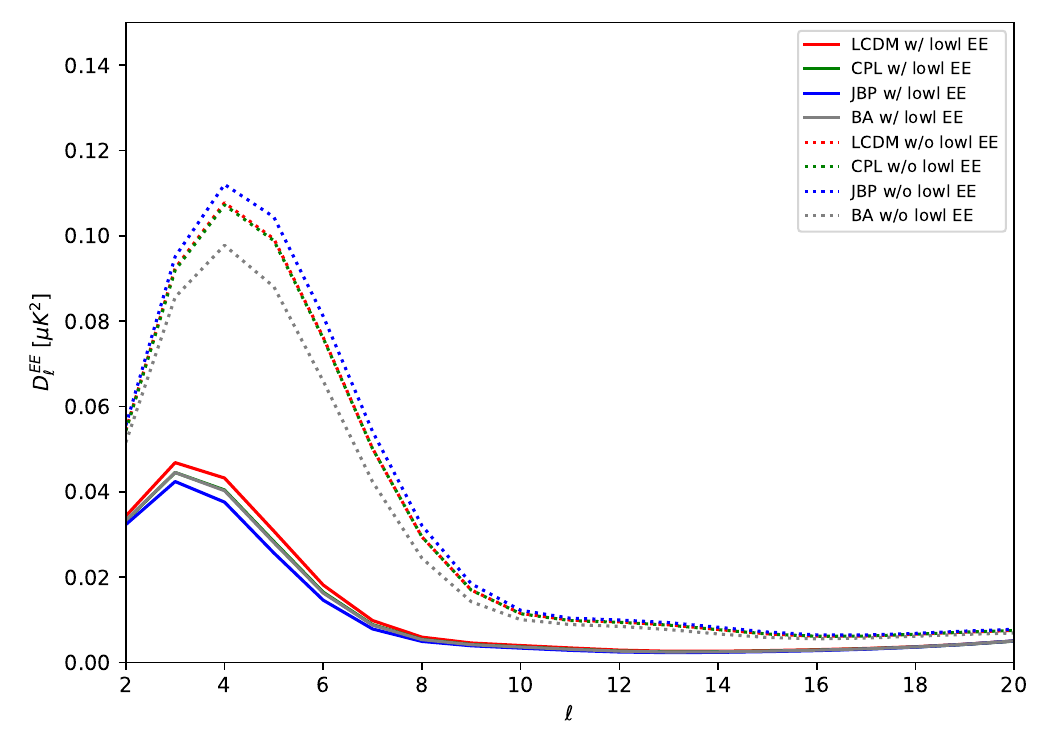}}
    \subfloat[Planck\label{1b}]{
    \includegraphics[width=0.5\textwidth]{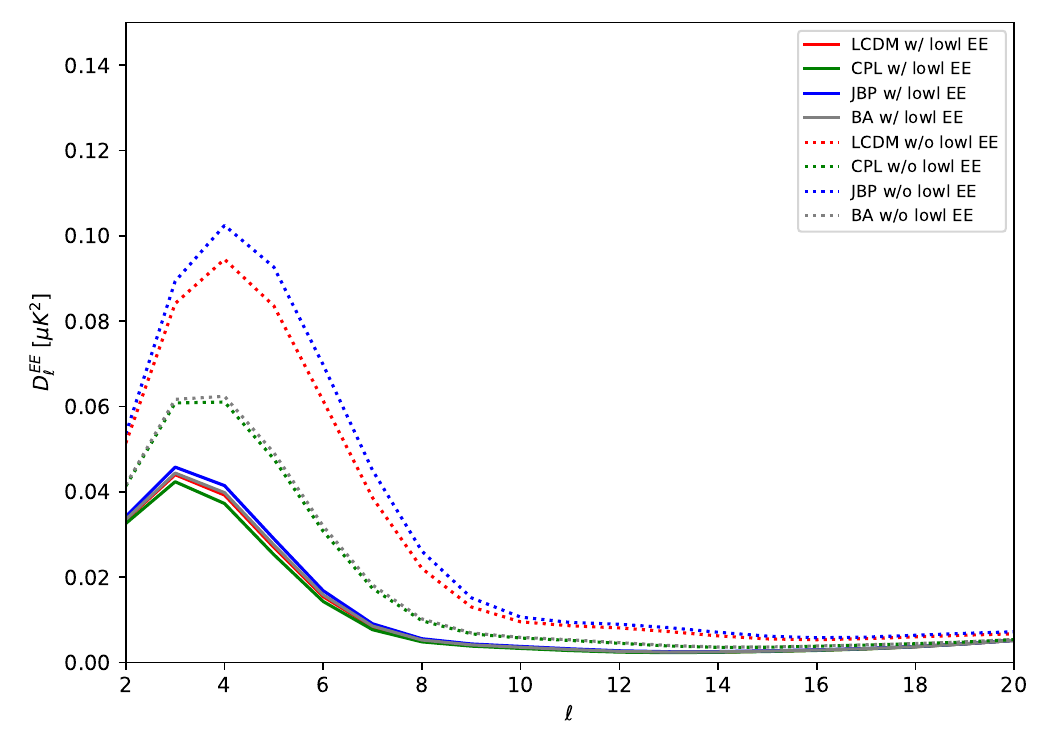}}
    \caption{\rthis{The best-fit theory low-$\ell$ EE power spectra for CMB$+$BAO$+$SNe, for each dark energy parameterization and CMB dataset, with and without low-$\ell$ EE data included in the fit}. Here, $D_{\ell}^{EE}$ is defined as $\frac{\ell(\ell+1)C_{\ell}^{EE}}{2\pi}T_\mathrm{CMB}^2$.}
    \label{fig1}
\end{figure}

For $\tau_\mathrm{reio}$, we find negligible shift across all comparisons except upon the exclusion of low-$\ell$ EE data, as expected. Excluding low-$\ell$ EE data shifts the central $\tau_\mathrm{reio}$ values upward and the posteriors become significantly wider. In certain cases, we get only a lower limit on $\tau_\mathrm{reio}$ when excluding low-$\ell$ EE data as the posterior constraints are now driven purely by the $A_s-\tau_\mathrm{reio}$ degeneracy rather than a direct polarization constraint on reionization. In Fig.~\ref{fig1}, we show the effect of removing the large-scale EE polarization data on the EE power spectrum. The reionization bump, which is directly constrained by low-$\ell$ EE data, shows an increase in the allowed signal amplitude from $\mathcal{O}(10^{-2}) \mu\mathrm{K}^2$ when including low-$\ell$ EE data to $\mathcal{O}(10^{-1}) \mu\mathrm{K}^2$ when excluding it, reflecting the loss of the direct polarization constraint~\cite{giare_2024}.

Comparing $\sigma_8$, we find $\lesssim 0.6 \sigma$ consistency when changing the CMB dataset (ACT or Planck) and between DE parameterizations. However, excluding the low-$\ell$ EE data shifts $\sigma_8$ by $\sim 1.5\sigma$ to higher values in $\Lambda$CDM (e.g., $0.8130 \pm 0.0048$ with low-$\ell$ EE and $0.8264^{+0.0081}_{-0.0051}$ without low-$\ell$ EE for ACT DR6). Exclusion of low-$\ell$ EE polarization data leads to a large relaxation in error bars ($\sim 40\%$). Hence, we can conclude that the shift is due to the central values shifting upward. For CPL, JBP, and BA parameterizations, no such shift is observed. Although excluding low-$\ell$ EE data does widen the $\sigma_8$ error bars in the DE parameterizations, the relaxation is small $(\sim 5-20 \%)$. The absence of a significant shift in $\sigma_8$ can be attributed to their intrinsically large $\sigma_8$ uncertainties due to the additional freedom in $w_0$ and $w_a$, which absorbs any upward pull on the central value. Excluding the large-scale EE polarization data does not significantly shift ($\lesssim 0.5\sigma$) the DE parameter values - $w_0$ and $w_a$.

We next consider the joint constraints on the DE EOS parameters and their sensitivity to low-$\ell$ EE data. \rthis{For CPL (Fig.~\ref{2a}), JBP (Fig.~\ref{3a}), and BA (Fig.~\ref{4a}) parameterizations, the Planck$+$BAO$+$SNe dataset yields joint $(w_0, w_a)$ constraints consistent with $(-1, 0)$ within $2\sigma$ when excluding low-$\ell$ EE data, while for ACT DR6, the joint contours encompass $w_a=0$ but not $w_0=-1$ within $2\sigma$ for CPL (Fig.~\ref{2b}) and BA (Fig.~\ref{4b}) only. For JBP in the ACT DR6 case, the consistency with $(w_0, w_a)=(-1, 0)$ is within $2\sigma$ when excluding low-$\ell$ EE data (Fig.~\ref{3b}). Broadly, we find that in all cases, excluding low-$\ell$ EE polarization data makes the joint $(w_0, w_a)$ contours consistent with the standard $\Lambda$CDM value of $(-1, 0)$.}

\begin{figure}[H]
    \centering
    \subfloat[Planck\label{2a}]{\includegraphics[width=0.3\textwidth,keepaspectratio]{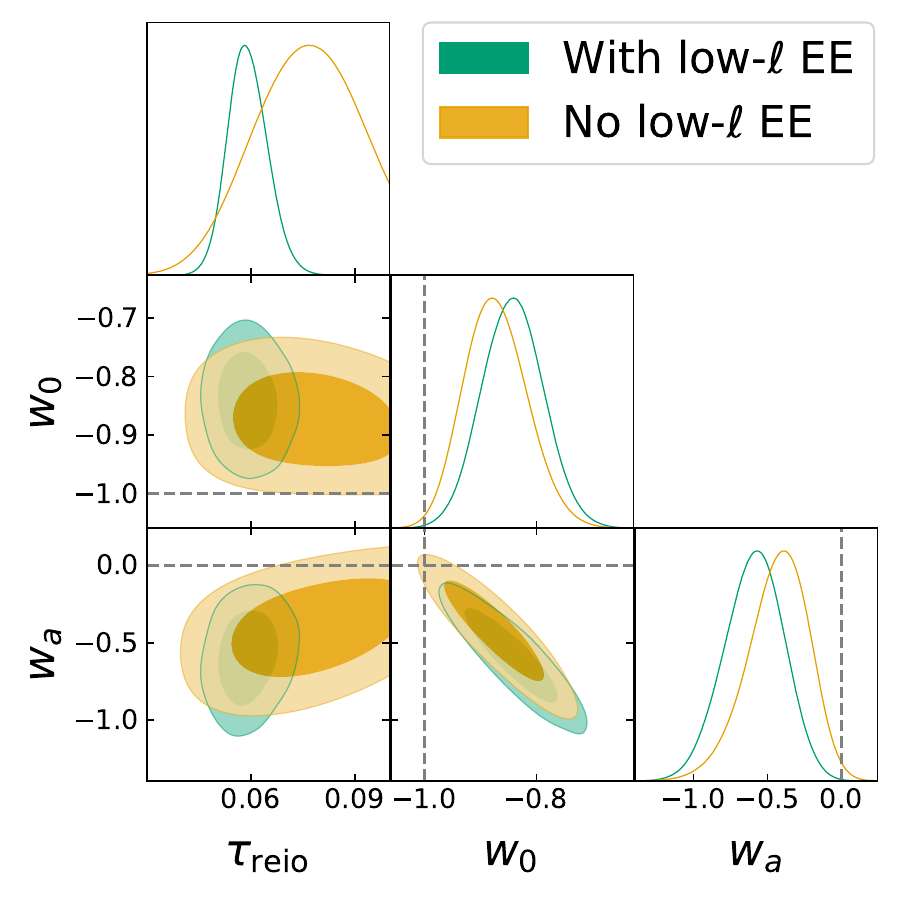}}
    \subfloat[ACT DR6\label{2b}]{\includegraphics[width=0.3\textwidth,keepaspectratio]{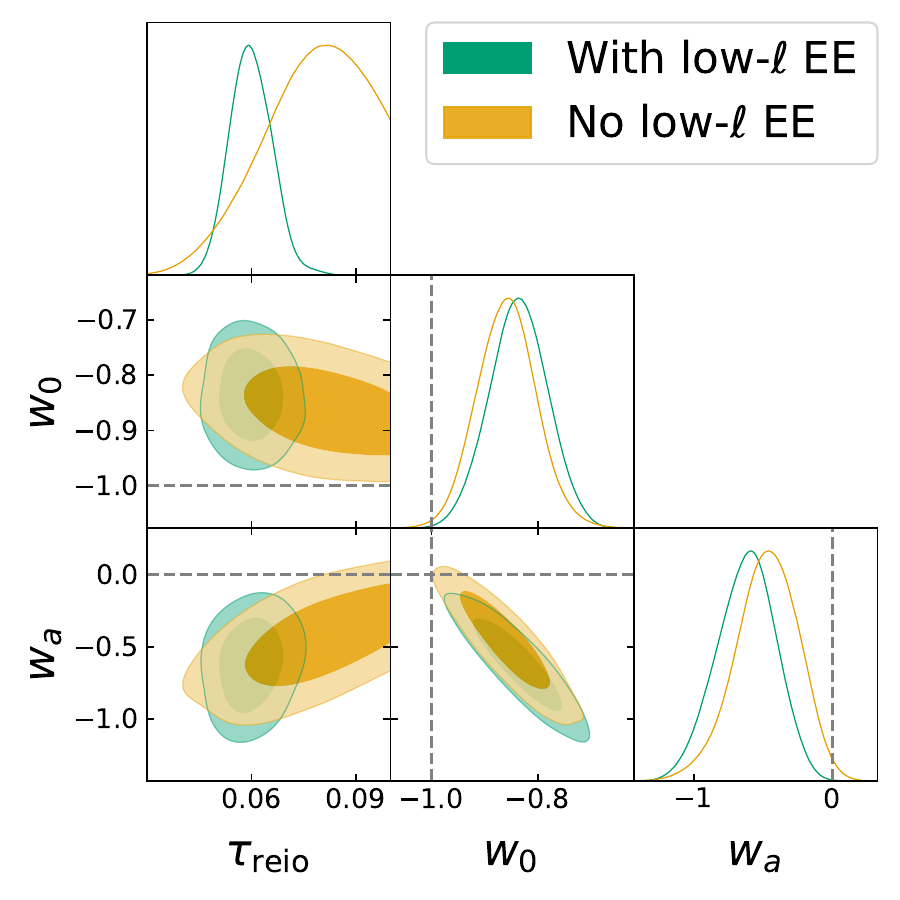}}
    \subfloat[EOS\label{2c}]{\includegraphics[width=0.4\textwidth,keepaspectratio]{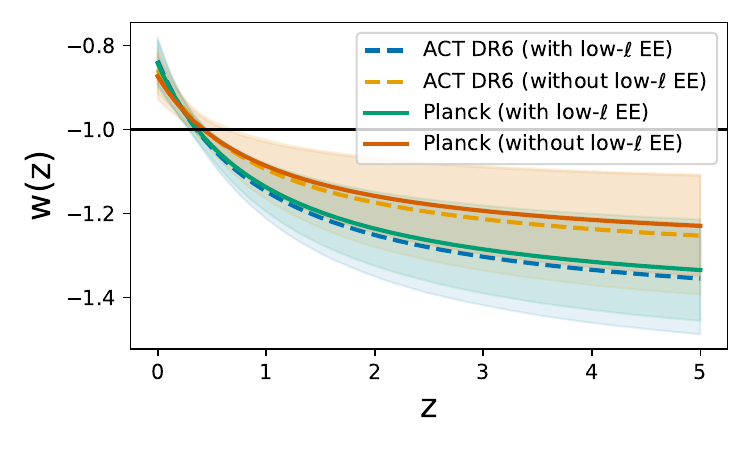}} \\
    \caption{$68\%$ and $95\%$ contour plots for CMB$+$BAO$+$SNe considering the CPL parameterization.}
    \label{fig2}
\end{figure}

\begin{figure}[H]
    \centering
    \subfloat[Planck\label{3a}]{\includegraphics[width=0.3\textwidth,keepaspectratio]{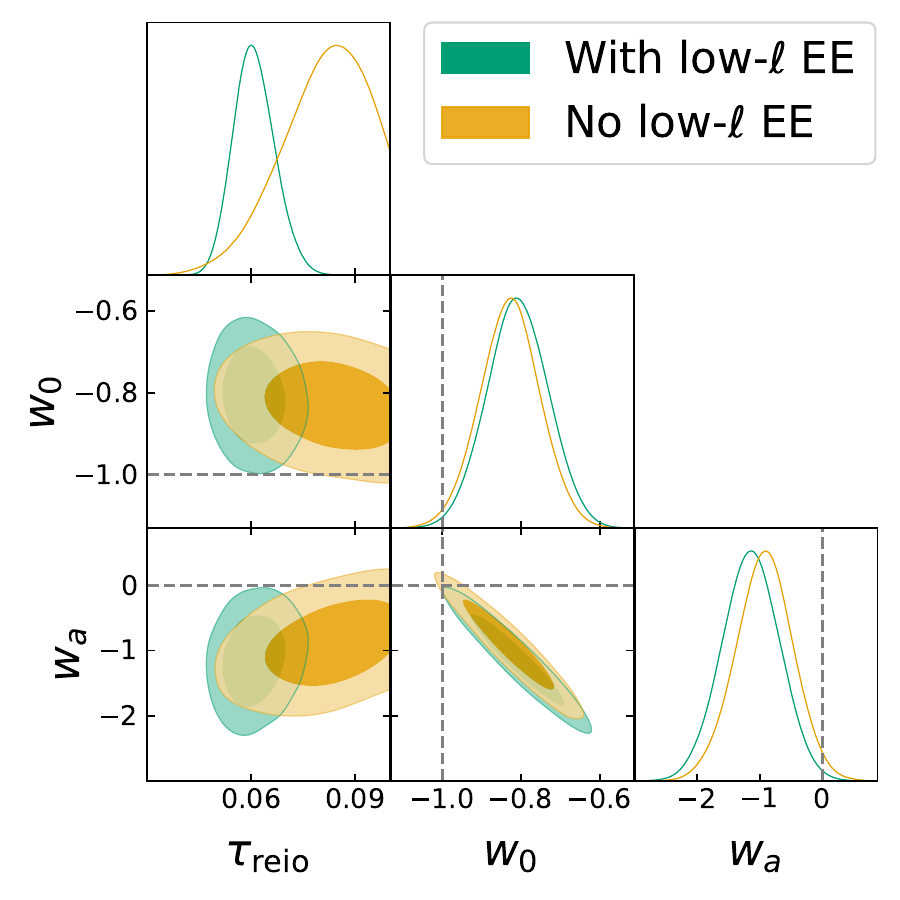}}
    \subfloat[ACT DR6\label{3b}]{\includegraphics[width=0.3\textwidth,keepaspectratio]{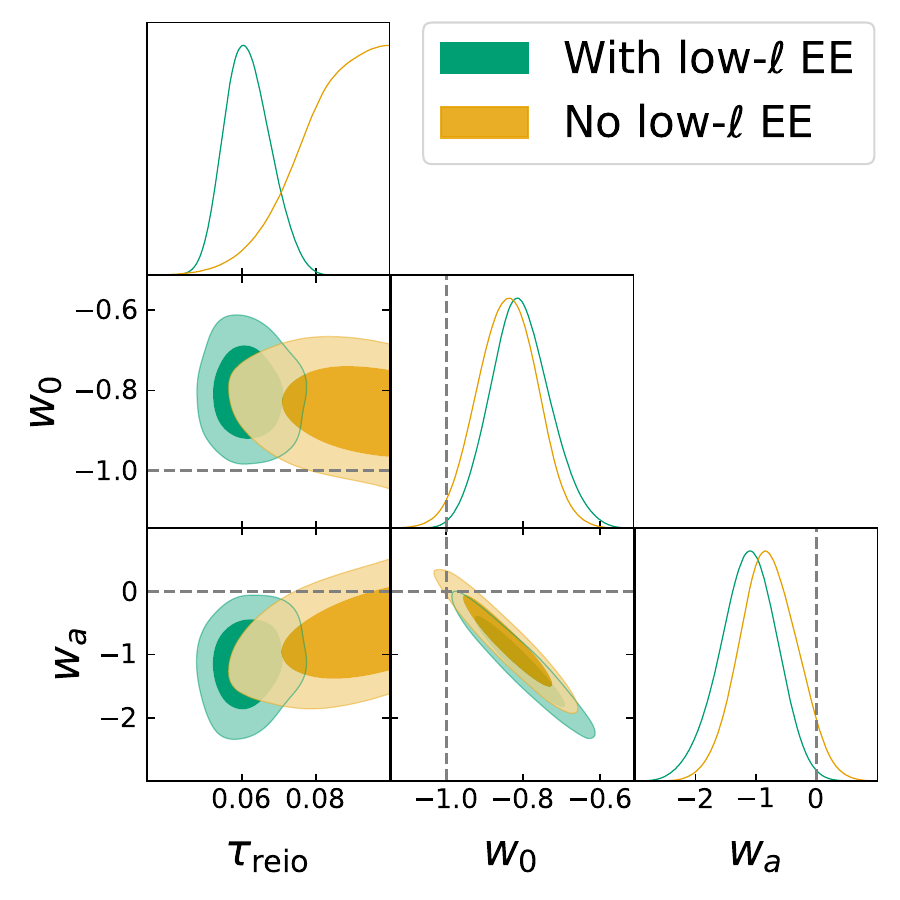}}
    \subfloat[EOS\label{3c}]{\includegraphics[width=0.4\textwidth,keepaspectratio]{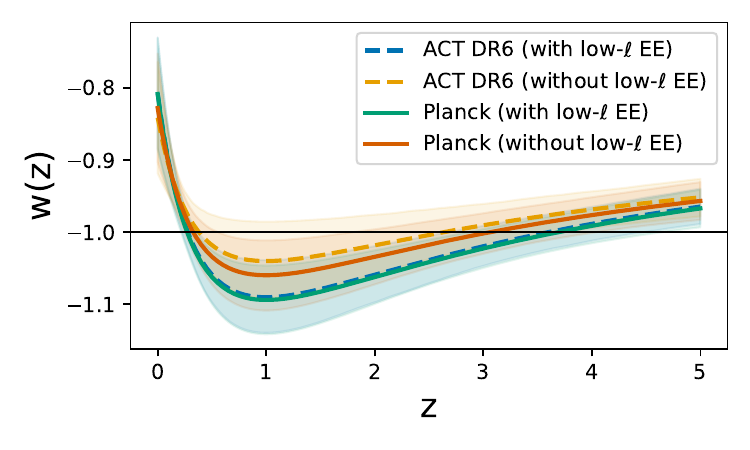}} \\
    \caption{$68\%$ and $95\%$ credible intervals for CMB$+$BAO$+$SNe considering the JBP parameterization.}
    \label{fig3}
\end{figure}

\begin{figure}[H]
    \centering
    \subfloat[Planck\label{4a}]{\includegraphics[width=0.3\textwidth,keepaspectratio]{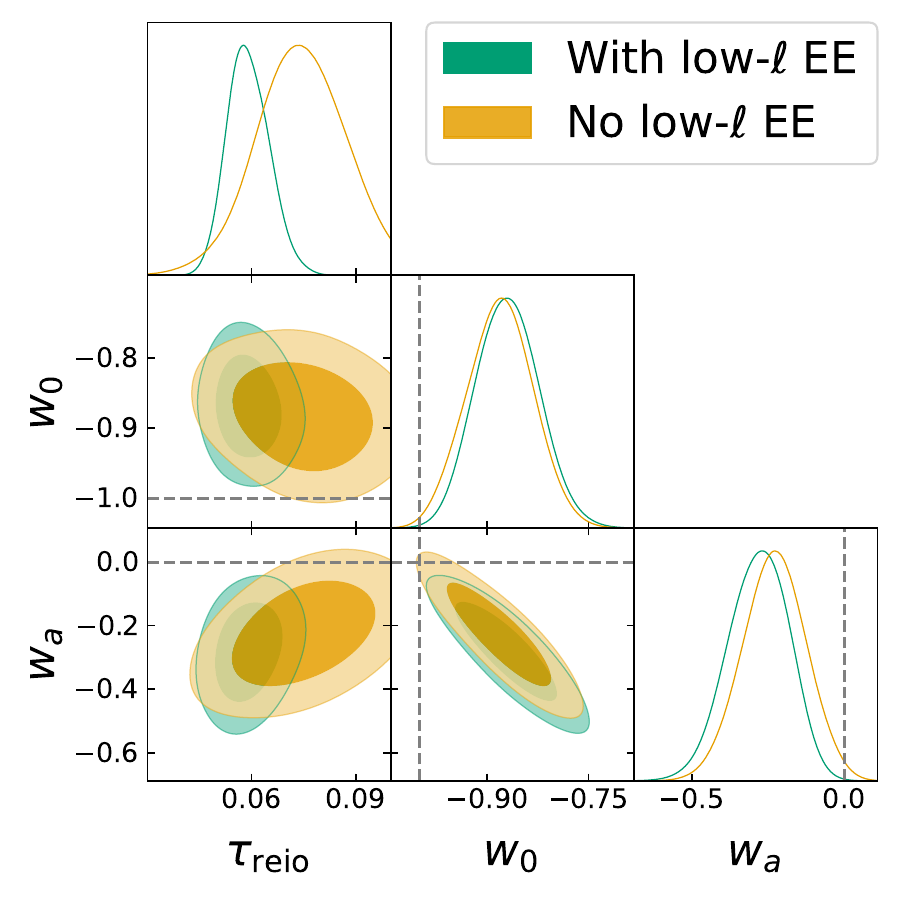}}
    \subfloat[ACT DR6\label{4b}]{\includegraphics[width=0.3\textwidth,keepaspectratio]{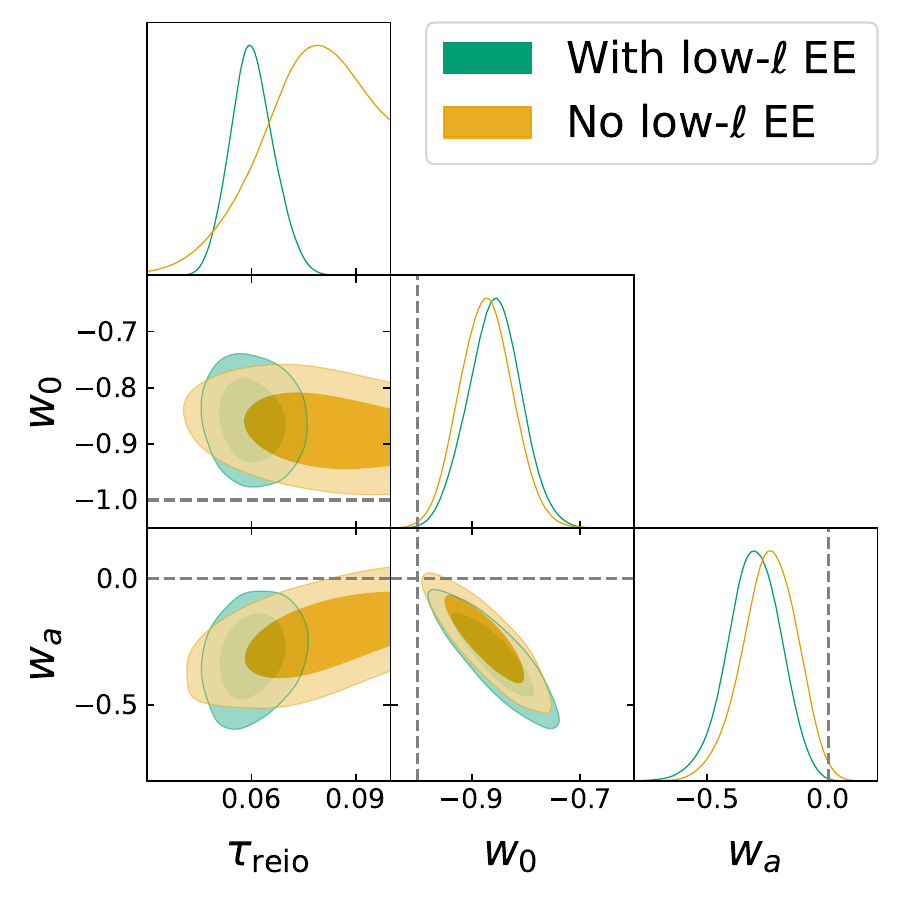}}
    \subfloat[EOS\label{4c}]{\includegraphics[width=0.4\textwidth,keepaspectratio]{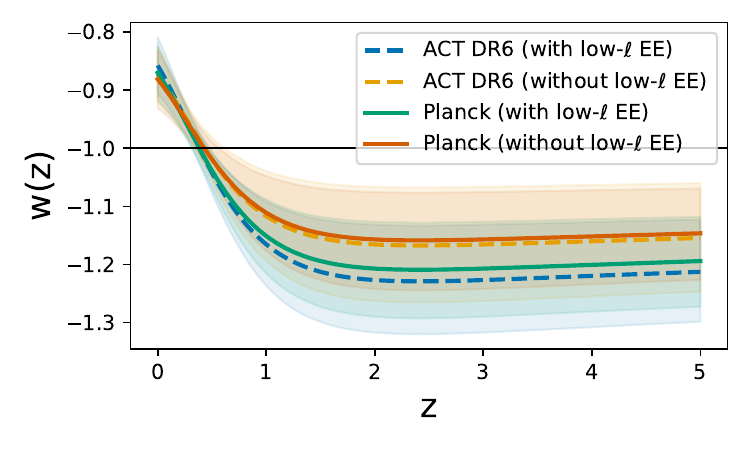}} \\
    \caption{$68\%$ and $95\%$ credible intervals for CMB$+$BAO$+$SNe considering the BA parameterization.}
    \label{fig4}
\end{figure}

The weighted Pearson correlation matrices (Fig.~\ref{fig5}) provide further insights into the parameter degeneracies. In $\Lambda$CDM, the $A_s-\tau_\mathrm{reio}$ correlation strengthens upon exclusion of low-$\ell$ EE data, reflecting the unbroken $A_se^{-2\tau_\mathrm{reio}}$ degeneracy. This propagates into $\sigma_8$ with the $A_s-\sigma_8$ and $\tau_\mathrm{reio}-\sigma_8$ correlations being strengthened, consistent with the $\sim 1.5 \sigma$ shift in $\sigma_8$ as discussed above. Excluding low-$\ell$ EE data strengthens the negative $\Omega_ch^2-\tau_\mathrm{reio}$ correlation from weak to moderate, whereas the $\Omega_ch^2-\sigma_8$ correlation decreases from weakly positive to negligible in ACT DR6 and to weakly negative in Planck. This suggests that the $A_s-\tau_\mathrm{reio}$ degeneracy becomes the dominant factor governing $\sigma_8$. \rthis{One can understand the upward shift of $\tau_\mathrm{reio}$ when low-$\ell$ EE polarization data is excluded through its degeneracy with $\Omega_m$. Fig. 1 of Ref.~\cite{sailer_2025} shows that lower values of $\Omega_m$ (as preferred by BAO data) require higher $\tau_\mathrm{reio}$ values. In our analysis, from Fig.~\ref{fig5}, we see that $\Omega_ch^2$ and $\tau_\mathrm{reio}$ are negatively correlated and this anti-correlation strengthens upon exclusion of low-$\ell$ EE data. We also find a positive correlation between $\Omega_bh^2$ and $\tau_\mathrm{reio}$ (not shown here). However, its magnitude is consistently smaller than that of the $\Omega_ch^2-\tau_\mathrm{reio}$ correlation across all parameterizations and datasets. We therefore conclude that the $\Omega_m-\tau_\mathrm{reio}$ degeneracy is primarily driven by $\Omega_ch^2$, with the decrease in $\Omega_m$ preferred by BAO being accommodated by the corresponding upward shift in $\tau_\mathrm{reio}$.}

\begin{figure}[H]
    \centering
    \subfloat[ACT DR6$+$BAO$+$SNe\label{5a}]{
    \includegraphics[width=\linewidth]{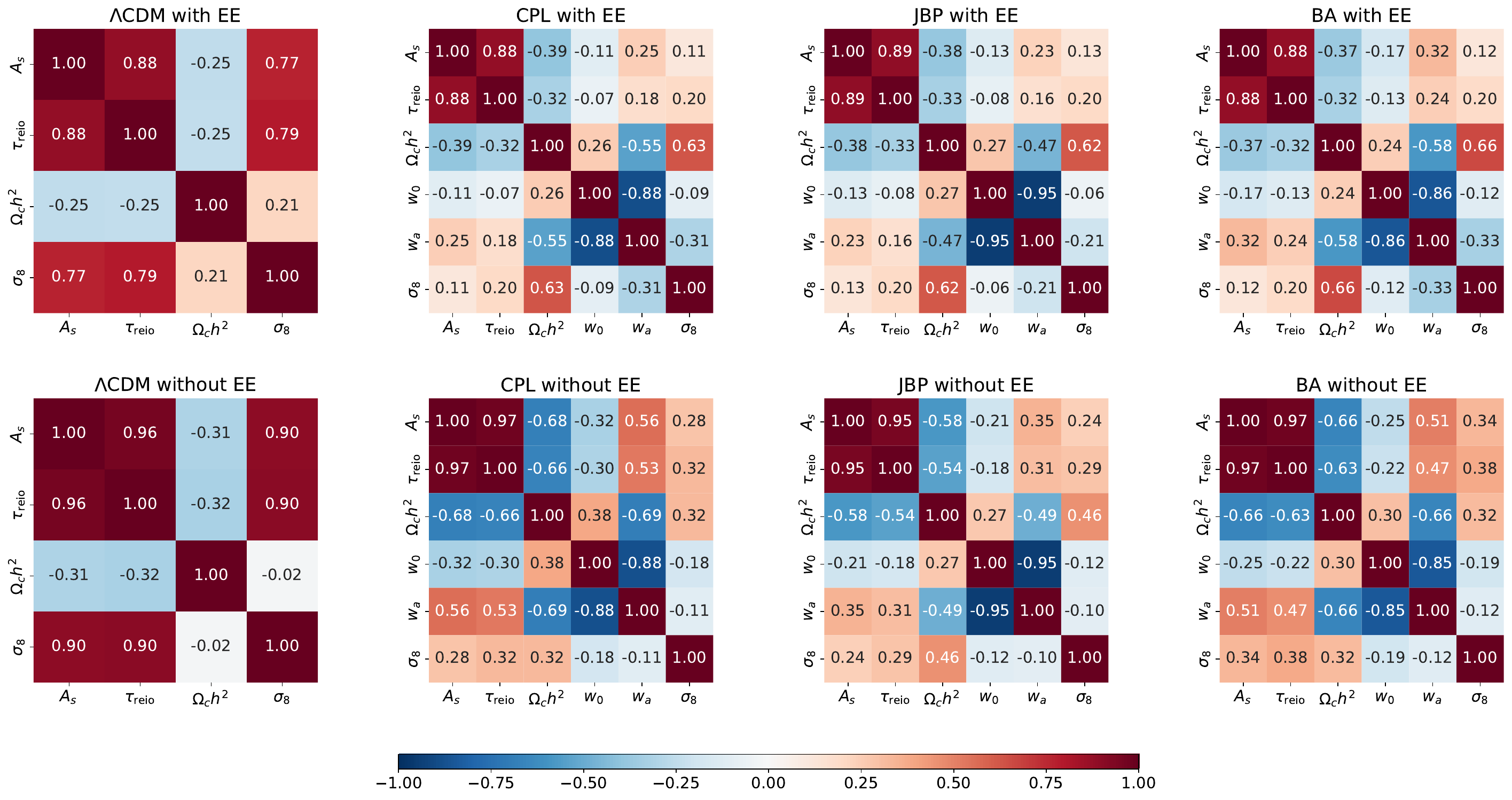}} \\
    \subfloat[Planck$+$BAO$+$SNe\label{5b}]{
    \includegraphics[width=\linewidth]{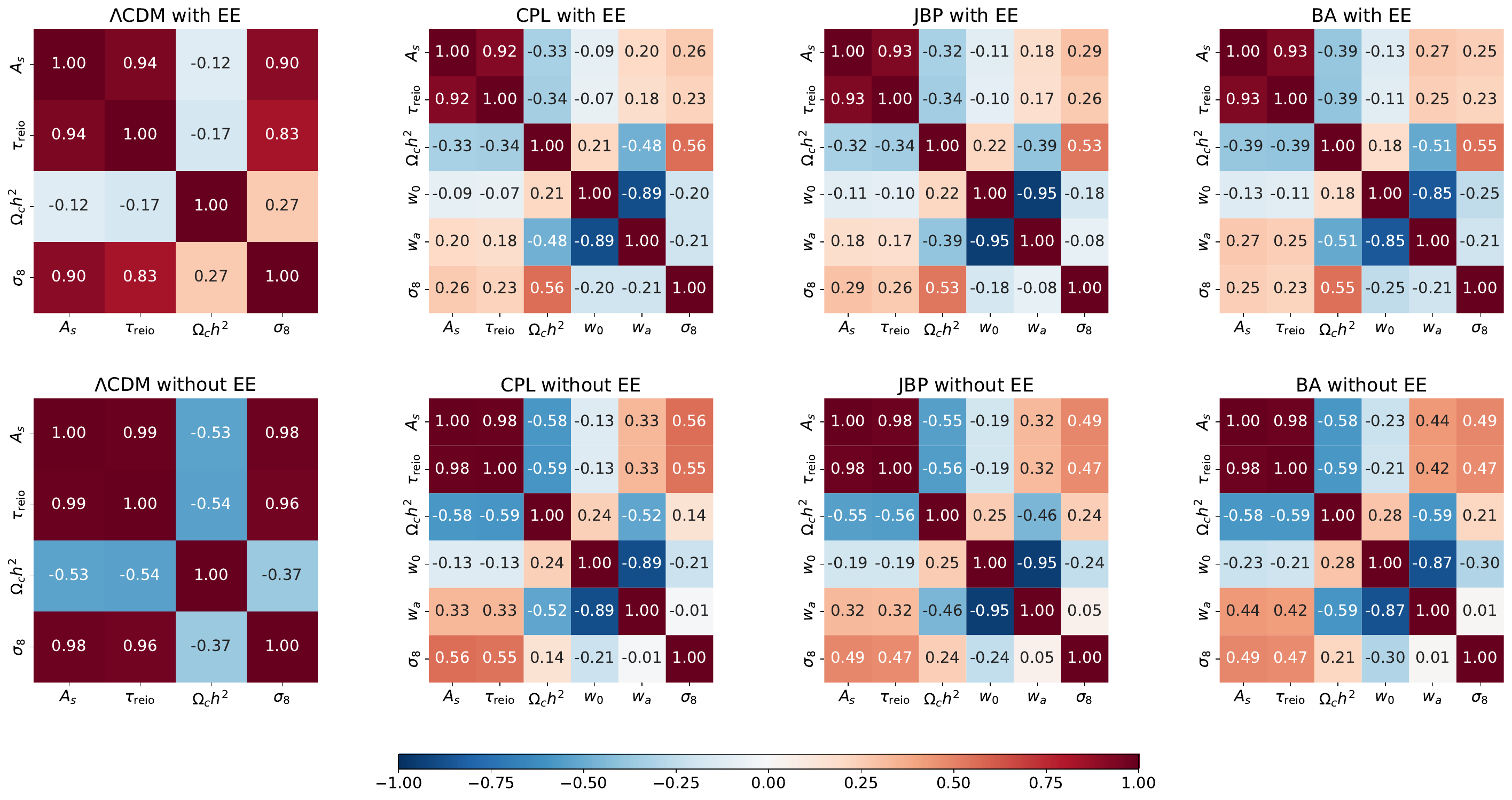}}
    \caption{Correlation among key model parameters for CMB$+$BAO$+$SNe.}
    \label{fig5}
\end{figure}

In contrast, for the DE parameterizations, the $A_s-\sigma_8$ and $\sigma_8-\tau_\mathrm{reio}$ correlations remain weak (for ACT DR6), and moderate (for Planck) regardless of the inclusion of low-$\ell$ EE \rthis{data}, showing only a mild strengthening upon its exclusion compared to $\Lambda$CDM. Additionally, in contrast to $\Lambda$CDM, there is  a strong correlation between $\Omega_ch^2$ and $\sigma_8$ when including large-scale EE polarization data, which decreases to moderate upon its exclusion. Hence, in the DE parameterizations studied in this work, $\sigma_8$ is affected by both the $A_s-\tau_\mathrm{reio}$ degeneracy as well as its correlation with $\Omega_ch^2$. This is consistent with the small $\sigma_8$ error bar relaxation observed in these parameterizations discussed above. The $w_0-\tau_\mathrm{reio}$ correlation remains weakly negative regardless of low-$\ell$ EE inclusion, whereas the $w_a-\tau_\mathrm{reio}$ correlation is weak when  low-$\ell$ EE data are included, and strengthens to a moderate  level when they are excluded. Both $w_0$ and $w_a$ show weak anti-correlation with $\sigma_8$, unaffected by the inclusion or exclusion of low-$\ell$ EE data.

The weak $A_s-\sigma_8$ correlation in the DE parameterizations can be understood as a consequence of the dependence of $\sigma_8$ on the growth factor $D(z, w_0, w_a)$. In DE parameterizations, the growth factor acquires additional dependence on $w_0$ and $w_a$, beyond the usual $\Omega_m$ and $H_0$ dependence. This allows part of the correlation to be absorbed by the DE parameters in comparison to $\Lambda$CDM where the growth factor is dependent only on $\Omega_m$ and $H_0$.

For all three DE parameterizations, the $w_0-\tau_\mathrm{reio}$ and $w_0-A_s$ correlations are weakly negative while the $w_a-\tau_\mathrm{reio}$ and $w_a-A_s$ correlations are weakly positive when including low-$\ell$ EE data. Upon exclusion of low-$\ell$ EE data, these correlations strengthen - particularly $w_a-\tau_\mathrm{reio}$ and $w_a-A_s$,  which increase from weak to moderate. $w_0$ correlations also increase but they remain weaker than $w_a$. The opposing signs of the $w_0$ and $w_a$ correlations with $\tau_\mathrm{reio}$ and $A_s$ reflect the strong negative $w_0-w_a$ degeneracy, leading to partial cancellations of the net EOS response to the $\tau_\mathrm{reio}$ shift (Appendix~\ref{appA}). This is consistent with the small shifts observed in $w(z)$ upon exclusion of low-$\ell$ EE data. Higher values of $\tau_\mathrm{reio}$, obtained upon excluding low-$\ell$ EE data,  shift $w(z)$ towards the quintessence regime, with the  $1\sigma$ error band for the JBP parameterization lying entirely in the quintessence region as shown in Fig.~\ref{3c}. \rthis{As discussed in Section~\ref{sec:2}, the redshift dependence of the $w_a$ contribution to $w(z)$ differs across the three parameterizations. While the partial cancellation between the $w_0$ and $w_a$ responses to the $\tau_\mathrm{reio}$ shift (Appendix~\ref{appA}) occurs for all three parameterizations, the magnitude of this pull at intermediate redshifts is largest for JBP, owing to its particular functional form $f(z)=\frac{z}{(1+z)^2}$, which peaks at intermediate-$z$ before vanishing at high-$z$, so that $w(z) \rightarrow w_0$ asymptotically. This more pronounced intermediate redshift pull is responsible for the JBP EOS lying entirely within the quintessence region at $1\sigma$ (Fig.~\ref{3c}).} 

\begin{figure}[H]
    \centering
    \includegraphics[width=\linewidth]{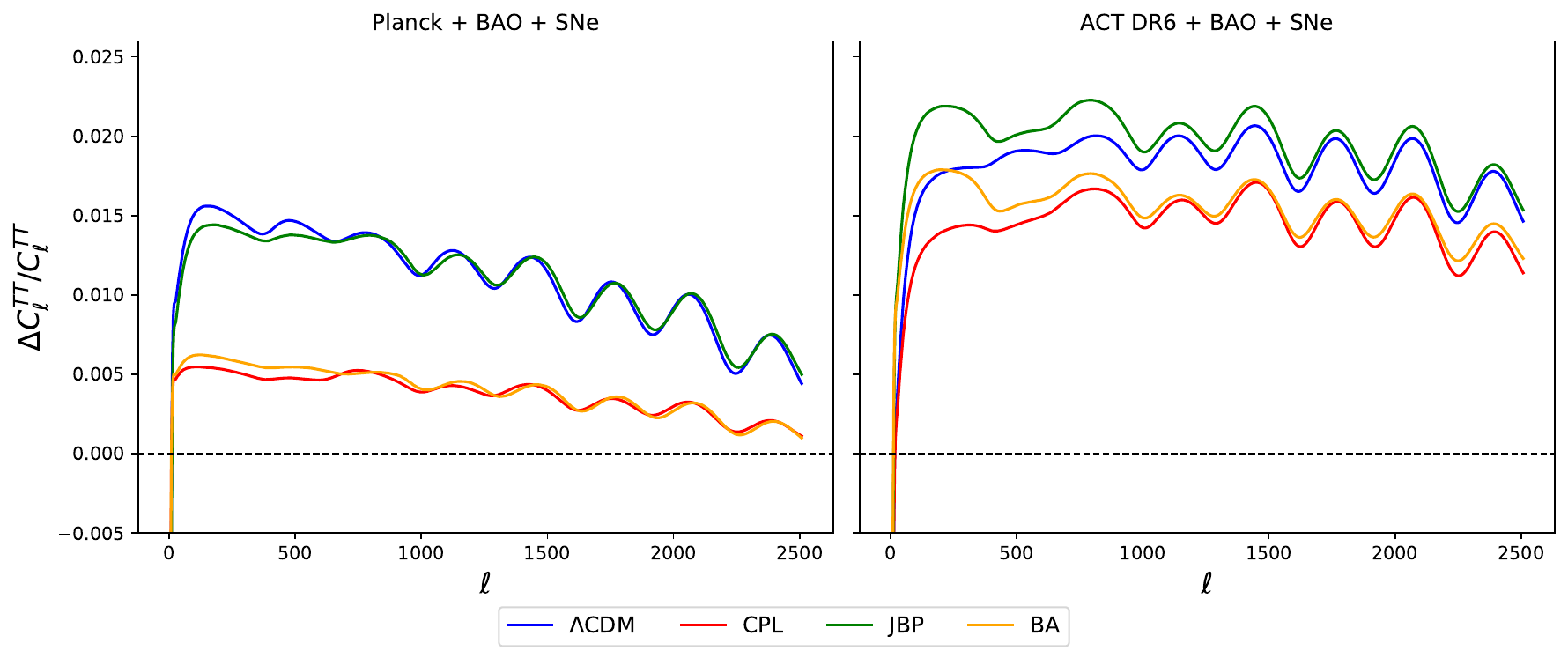}
    \caption{Relative shift of the TT power spectra when excluding low-$\ell$ EE data, using the spectrum with low-$\ell$ EE data as reference. ($\Delta C_\ell^{TT} = C_\ell^{TT}(\text{incl. low-}{\ell}\ \text{EE}) 
- C_\ell^{TT}(\text{excl. low-}{\ell}\ \text{EE})$)}
    \label{fig6}
\end{figure}

The $A_s$ shifts, along with the corresponding $\tau_\mathrm{reio}$ shifts through the $A_se^{-2\tau_\mathrm{reio}}$ degeneracy, are reflected in the relative shift of the TT power spectrum (cf. Fig.~\ref{fig6}), with $\Lambda$CDM and JBP showing larger $C_\ell^{TT}$ shifts consistent with their larger $A_s$ tensions, while CPL and BA show smaller shifts. For $\tau_\mathrm{reio}$, a quantitative tension estimate is not possible in all cases, though the upward shift in central values when excluding low-$\ell$ EE data is consistent with the observed TT amplitude increase.

Next, we compare the different parameterizations based on $\Delta$AIC and $\Delta$DIC values (Fig.~\ref{fig7}).

\subsection{Model Comparison}
\label{aic_dic}

\begin{figure}[H]
    \centering
    \includegraphics[width=\linewidth]{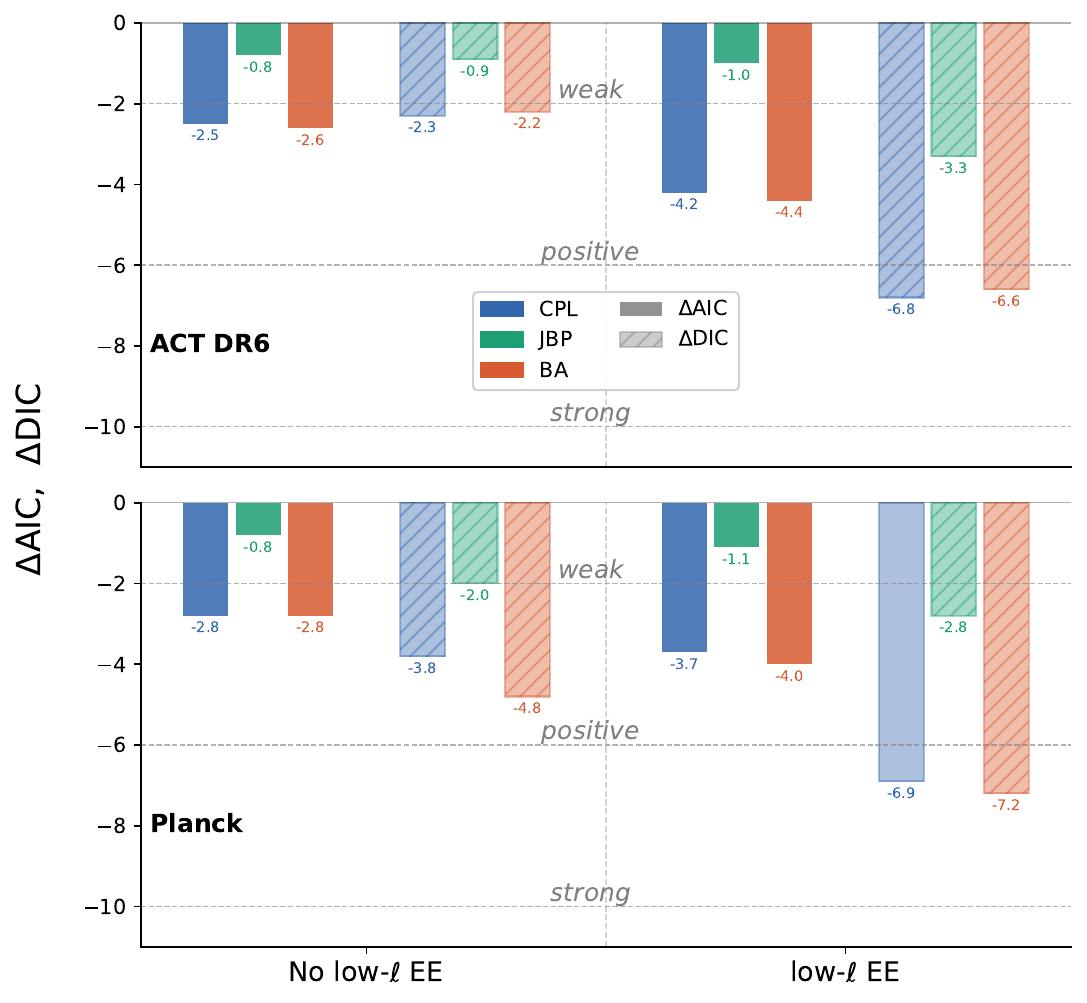}
    \caption{Model comparison results. When $-2\leq\Delta\mathrm{AIC},\Delta\mathrm{DIC}<0$, the evidence is \textit{weak} in favor of the model, for $-6\leq\Delta\mathrm{AIC},\Delta\mathrm{DIC}<-2$, the evidence is \textit{positive}, for $-10\leq\Delta\mathrm{AIC},\Delta\mathrm{DIC}<-6$, there is \textit{strong} evidence while for $\Delta\mathrm{AIC},\Delta\mathrm{DIC}<-10$ , the evidence in favor of the model under consideration is \textit{very strong}. The top and bottom panels show $\Delta$AIC and $\Delta$DIC for ACT DR6 and Planck CMB dataset combinations, respectively. All comparisons are performed relative to $\Lambda$CDM, which is taken as the reference (null) model (Eqns.~\ref{eqn5} and~\ref{eqn6}).}
    \label{fig7}
\end{figure}

 The $\Delta$AIC and $\Delta$DIC values are shown in Fig.~\ref{fig7}.  For the ACT DR6 CMB$+$BAO$+$SNe dataset, we find positive evidence in favor of the CPL and BA parameterizations and weak evidence in favor of JBP over $\Lambda$CDM. This is robust to the inclusion or exclusion of low-$\ell$ EE data. The same evidence strength is found when replacing ACT DR6 CMB data with Planck CMB dataset.

Comparing models based on $\Delta$DIC, we find that CPL and BA show positive evidence over $\Lambda$CDM when excluding low-$\ell$ EE data, changing to strong evidence upon its inclusion, for both Planck and ACT DR6 CMB datasets. \rthis{However, for both ACT DR6 and Planck CMB datasets, JBP shows weak and positive evidence over $\Lambda$CDM when excluding and including low-$\ell$ EE data, respectively.}

\begin{figure}[H]
    \centering
    \includegraphics[width=\linewidth]{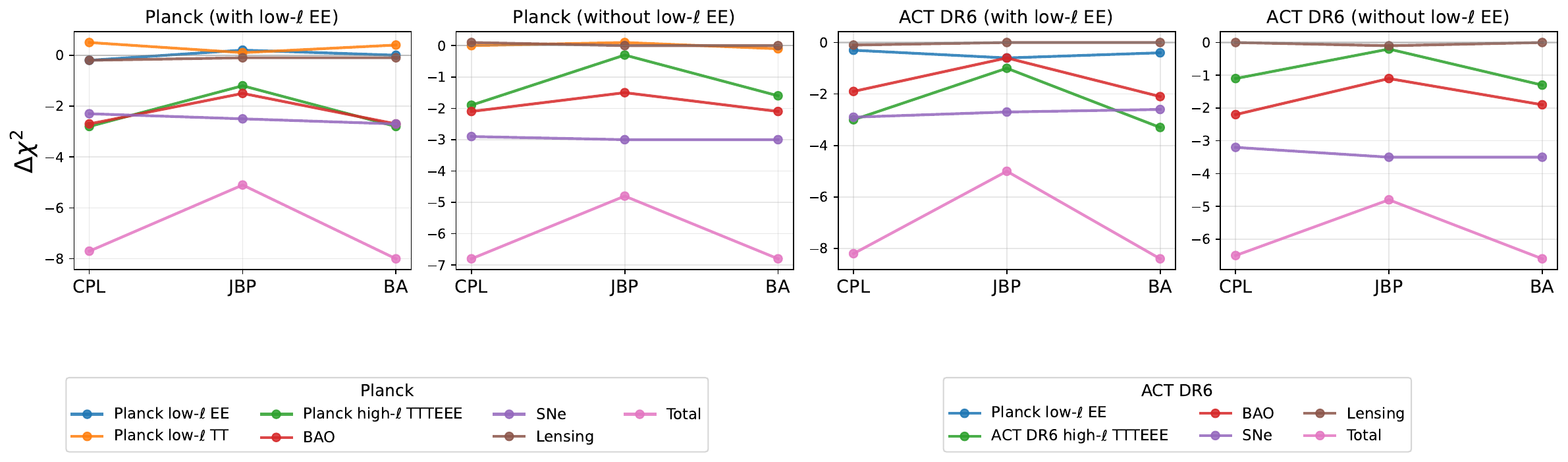}
    \caption{Per-likelihood contribution and total likelihood for CMB$+$BAO$+$SNe data where $\Delta \chi^2 = \chi^2$($\mathcal{M}$) - $\chi^2$($\Lambda$CDM). $\mathcal{M}$ refers to CPL, JBP or BA models.}
    \label{fig8}
\end{figure}

We find that SNe likelihood is the primary driver of $\Delta\chi^2$ improvement across all DE parameterizations (Fig.~\ref{fig8}). The high-$\ell$ TTTEEE CMB data provide a non-negligible contribution for CPL and BA but less so for JBP. BAO contributes a consistent improvement of $\Delta \chi^2 \sim -2 $ across all cases. The contributions from low-$\ell$ EE and lensing datasets are negligible in comparison. Since DE parameterizations have two extra parameters, AIC requires $\Delta \chi^2 < -6$ to show positive evidence over $\Lambda$CDM. This threshold is met by CPL and BA in all cases but not consistently by JBP. These findings are robust to the choice of the CMB dataset - the only difference in the Planck case is the additional low-$\ell$ TT likelihood which contributes negligibly to $\Delta \chi^2$ and does not affect conclusions regarding model comparison.

\section{Conclusions}
\label{sec:5}

In this work, we studied the effect of large-scale EE CMB polarization data, which constrain $\tau_\mathrm{reio}$ tightly, for three different DE parameterizations - CPL, JBP,  and BA. We analyzed how correlations between the different parameters affect the EOS $w(z)$ and also performed model comparison based on AIC and DIC. 

We find an anti-correlation between $A_s$ and $\Omega_ch^2$, which gets enhanced when excluding low-$\ell$ EE data, with the effect being more prominent in CPL, JBP,  and BA compared to $\Lambda$CDM. Constraints on $H_0$ remain robust across all cases, whereas $n_s$ and $\Omega_bh^2$ are mildly sensitive - $(1-1.2)\sigma$ and $(1.3-1.8)\sigma$, respectively - to the CMB dataset used. They are insensitive to the inclusion of low-$\ell$ EE data and the DE parameterization used. We notice an upward $(1.4-1.8) \sigma$ shift in $A_s$ for $\Lambda$CDM and JBP and a mild $\sim 1\sigma$ shift in BA and CPL when excluding low-$\ell$ EE data. This shift is consistent with the fact that excluding low-$\ell$ EE data produces a smaller relaxation in $A_s$ error bars for $\Lambda$CDM and JBP relative to CPL and BA. There exists a strong correlation between $A_s$ and $\tau_\mathrm{reio}$ due to the $A_se^{-2\tau_\mathrm{reio}}$ degeneracy at smaller scales. This correlation strengthens when excluding low-$\ell$ EE data, which is primarily responsible for constraining $\tau_\mathrm{reio}$. The value of $\tau_\mathrm{reio}$  increases and due to its correlation with $A_s$, $A_s$ values increase accordingly. In $\Lambda$CDM, the $\sigma_8$ values shift upward by $1.5\sigma$ as a consequence of its strong correlation with both $A_s$ and $\tau_\mathrm{reio}$. This is driven by a genuine shift in central $\sigma_8$ values when excluding low-$\ell$ EE data. However, the story is different for CPL, JBP and BA parameterizations. There is a strengthening of the $\sigma_8$ correlations with $A_s$ and $\tau_\mathrm{reio}$, but not as prominently as in $\Lambda$CDM. Combined with the fact that $\Omega_ch^2$ also shows a non-negligible correlation with $\sigma_8$, this explains the observation that the error bars in CPL, JBP and BA do not relax as much when excluding low-$\ell$ EE data and the additional freedom provided by $w_0$ and $w_a$.     
The correlations of $A_s$ and $\tau_\mathrm{reio}$ with $w_0$ are weakly negative, while their correlations with $w_a$ are weakly positive when low-$\ell$ EE data are included \footnote{The actual strength depends on which CMB data is used.}. When these data are excluded, the correlations with  $w_0$  increase slightly, but the dominant effect arises from the correlations of  $w_a$ with  $A_s$ and $\tau_\mathrm{reio}$. The net effect is a slight shift towards the quintessence region in all three DE parameterizations. Most notably, the EOS curve for the JBP parameterization lies in the quintessence region within the $1\sigma$ error band. \rthis{Therefore, the tendency of the EOS to shift towards the quintessence regime upon low-$\ell$ EE polarization data exclusion is not unique to the CPL parameterization and appears to be common to the DE parameterizations considered in this work, although the magnitude of the shift depends on the specific functional form of $w(z)$.}

We perform model comparison using AIC and DIC, where we consider the $\Lambda$CDM model as the reference model.
Considering AIC, we find positive evidence in favor of CPL and BA parameterizations and weak evidence in favor of JBP. This evidence is robust to the inclusion of low-$\ell$ EE data. DIC model comparison shows mild sensitivity to the dataset used. CPL and BA show positive evidence over $\Lambda$CDM when excluding low-$\ell$ EE data and strong evidence when including it for both Planck and ACT CMB datasets. JBP shows weak (positive) evidence over $\Lambda$CDM when excluding (including) low-$\ell$ EE data \rthis{for both CMB datasets}. The improvement in the fit for CPL, JBP and BA is driven mainly by the SNe likelihood, with additional contributions from high-$\ell$ TTTEEE for CPL and BA, a consistent $\Delta \chi^2 \sim -2$ from BAO and negligible impact from low-$\ell$ EE and lensing likelihoods. 

We note that while this work was in progress, the updated DES-Dovekie dataset was released~\cite{de_dovekie_2026}, which reanalyzed the DESY5 SNe sample using an improved photometric cross-calibration, white dwarf observations to cross-calibrate between DES and low redshift surveys and fixed a numerical approximation in the host galaxy color law. Using this updated sample, it was found that the evidence for DDE reduced from $4.2\sigma$ to $3.2\sigma$ which is closer to the evidence when using PP ($2.8\sigma$). Further, the constraint on $\Omega_m$ in a flat $\Lambda$CDM framework was found to be $0.330\pm0.015$ similar to the PP constraint - $0.334\pm0.018$. In light of this, we expect our results to be similar even when replacing the PP SNe dataset with the DES-Dovekie DESY5 dataset. The old DESY5 SNe data gave an $\Omega_m$ constraint of $0.352\pm0.017$ \cite{abbott_2024}. It is, therefore, clear that the results will depend on the SNe sample used, which we show in Appendix~\ref{appC}.

Refs.~\cite{dai_2026, jhaveri_2025, sullivan_2026} showed that the apparent preference for negative neutrino masses from CMB$+$BAO data is sensitive to $\tau_\mathrm{reio}$ and can be alleviated by a larger optical depth, largely independent of the details of the reionization history. Further, in Ref.~\cite{allali_2025}, it was found that the Hubble tension is reduced in $\Lambda$CDM, Early Dark Energy, and Dark radiation frameworks, when the low-$\ell$ EE polarization measurements were excluded due to correlations between $H_0$ and $\tau_\mathrm{reio}$. Our findings, alongside \cite{dai_2026, jhaveri_2025, allali_2025, giare_2024, upadhyay_2026}, suggest that $\tau_\mathrm{reio}$ is a common thread underlying multiple current cosmological tensions — from the preference for dynamical dark energy to the unphysical preference for negative neutrino masses. This further highlights the necessity to constrain $\tau_\mathrm{reio}$ more efficiently and not be dependent on only one dataset to constrain it.

\begin{acknowledgments}
SB would like to extend his gratitude to the University Grants Commission (UGC), Govt. of India for their continuous support through the Senior Research Fellowship, which has played a crucial role in the successful completion of our research. \rthis{ We also thank the anonymous referee for very constructive and useful comments on our manuscript.}
Computational work was supported by the National Supercomputing Mission (NSM), Government of India, through access to the ``PARAM SEVA'' facility at IIT Hyderabad. The NSM is implemented by the Centre for Development of Advanced Computing (C-DAC) with funding from the Ministry of Electronics and Information Technology (MeitY) and the Department of Science and Technology (DST). We also acknowledge the use of IUCAA HPC Computing facilities.
\end{acknowledgments}

\appendix

\section{EOS sensitivity to low-$\ell$ EE data}
\label{appA}

Here, we look at how the contributions from $w_0$ and $w_a$ explain the shifting of the EOS $w(z)$ towards the quintessence regime for the CPL, JBP and BA parameterizations. We define $\Delta w(z) = w(z)^{EE}-w(z)$, where $w(z)^{EE}$ represents the EOS when low-$\ell$ EE data is included while $w(z)$ is the EOS when low-$\ell$ EE data is excluded. This can also be seen as $\Delta w(z) = \Delta w_0 + \Delta w_a \cdot f(z)$, where $f(z)$ is the redshift scaling of the DE parameterization under consideration ($f(z) = \frac{z}{1+z}$ for CPL, $f(z) = \frac{z}{(1+z)^2}$ for JBP and $f(z)=\frac{z(1+z)}{1+z^2}$ for BA).

\begin{itemize}
    \item At z = 0, $\Delta w(z) > 0$. This means $w_0^{EE} > w_0$ since contribution from $w_a$ is 0. This shows the anti-correlation between $\tau_\mathrm{reio}$ and $w_0$. When $\tau_\mathrm{reio}$ increases, $w_0$ decreases, which is the case when low-$\ell$ EE data is removed.
    \item At intermediate redshifts, $f(z)$ contribution increases and the contribution from $w_a$ is no longer negligible. In this region $\Delta w_a$ is also negative. This implies $w_a^{EE}<w_a$. This means that as $\tau_\mathrm{reio}$ increases (no low-$\ell$ EE data), $w_a$ increases, showing the positive correlation between them. Further, we see that $\Delta w(z)$ is negative despite $\Delta w_0$ being positive. This means the contribution from $w_a$ outweighs the contribution from $w_0$. We also notice a crossing in all three parameterizations where $\Delta w_0 = -\Delta w_a \cdot f(z)$.
    \item At higher redshifts, the behaviour essentially depends on the form of $f(z)$. For CPL, $f(z) \rightarrow 1$ as $z \rightarrow \infty$, so the $w_a$ contribution saturates and $\Delta w(z)$ remains large and negative. For JBP, $f(z) \rightarrow 0$ as $z \rightarrow \infty$, allowing partial recovery of $\Delta w(z)$ towards $\Delta w_0$. BA behaves similarly to CPL.
\end{itemize}

The opposing signs of $\Delta w_0$ and $\Delta w_a$ observed across all three parameterizations are a direct consequence of the strong $w_0-w_a$ degeneracy. When $\tau_\mathrm{reio}$ shifts upwards due to exclusion of large-scale EE polarization constraints, $w_0$ and $w_a$ respond in opposite directions, leading to a partial cancellation in $\Delta w(z)$. Therefore, the net shift of $w(z)$ is also small.

\begin{figure}[H]
    \centering
    \includegraphics[width=\linewidth]{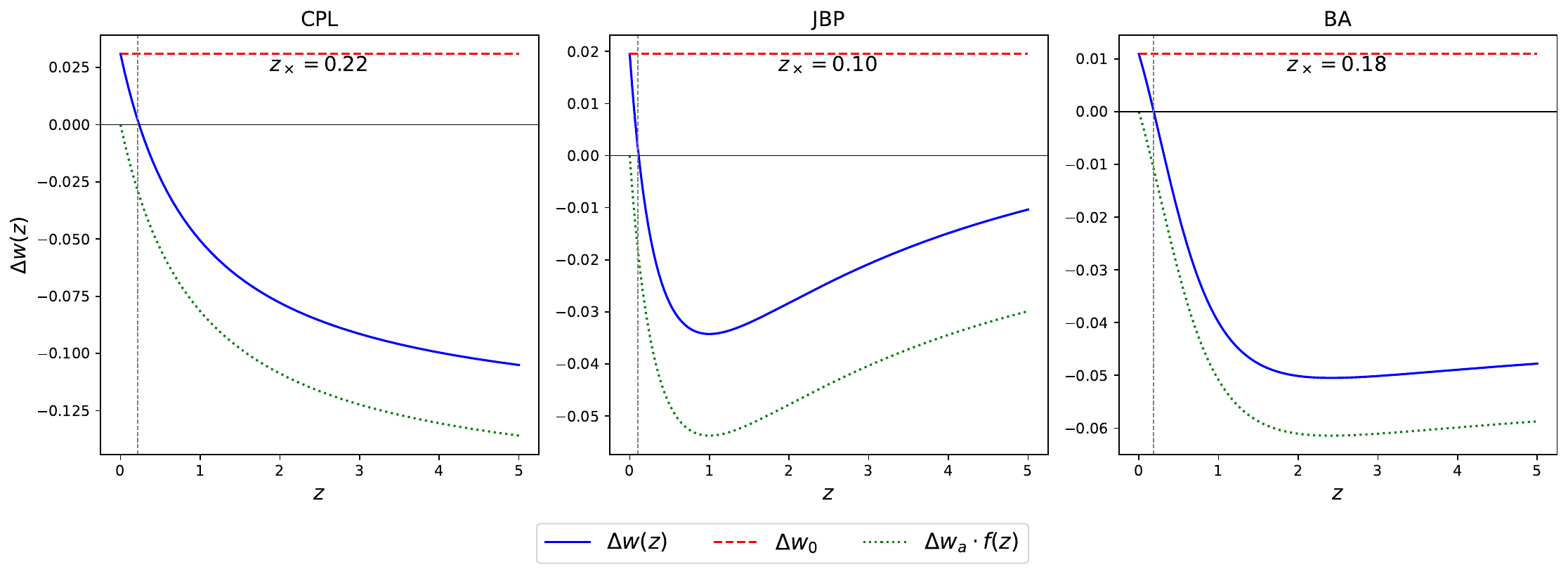}
    \caption{Cancellation of $w_0$ and $w_a$ contributions to $w(z)$ Planck CMB$+$BAO$+$PP. The crossing redshifts $z_\times = 0.22, 0.10, 0.18$ for CPL, JBP and BA respectively, mark the transition from $w_0$ dominated region to $w_a$ dominated region. The high redshift behaviour is determined by the form of $f(z)$ for each parameterization.}
    \label{fig9}
\end{figure}

\section{Likelihood contributions}
\label{appB}

Here, we present the total $\chi^2$ as well as the per-likelihood contributions to the total $\chi^2$ for all the dataset combinations considered in this work (Tables~\ref{table4} and~\ref{table5}). The per-likelihood $\Delta \chi^2$ contributions relative to $\Lambda$CDM can be visually seen in Fig.~\ref{fig8}.

\begin{table}[htbp!]
\caption{$\chi^2_\mathrm{min}$ contribution from each dataset for ACT DR6 CMB data. The first column for each parameterization is the bestfit $\chi^2$ values while the second column is the MAP $\chi^2$ values. Upper part of the table is for dataset combination including low-$\ell$ EE and lower part is excluding low-$\ell$ EE.}
\label{table4}
\centering
\resizebox{\textwidth}{!}{
   \begin{tabular}{l|c c|c c|c c|c c}
   \hline
   \thead{Dataset} & \multicolumn{2}{c|}{\thead{$\Lambda$CDM}} & \multicolumn{2}{c|}{\thead{CPL}} & \multicolumn{2}{c|}{\thead{JBP}} & \multicolumn{2}{c}{\thead{BA}} \\
   \hline
   \hline
   \rule{0pt}{2.5ex}planck\_2018\_lowl.EE\_sroll2 & $390.7$ & $390.8$ & $390.4$ & $390.4$ & $390.1$ & $390.6$ & $390.3$ & $390.9$ \\[0.5ex]
   act\_dr6\_cmbonly & $158.8$ & $159.3$ & $155.8$ & $155.9$ & $157.8$ & $156.9$ & $155.5$ & $155.6$ \\[0.5ex]
   bao.desi\_dr2.desi\_bao\_all & $11.9$ & $12.9$ & $10$ & $10.1$ & $11.3$ & $12.1$ & $9.8$ & $10$ \\ [0.5ex]
   sn.pantheonplus & $1405.9$ & $1405.6$ & $1403$ & $1403$ & $1403.2$ & $1403.4$ & $1403.3$ & $1402.9$ \\ [0.5ex]
   act\_dr6\_lenslike.ACTDR6LensLike & $19.7$ & $19.8$ & $19.6$ & $19.6$ & $19.7$ & $19.7$ & $19.7$ & $19.6$ \\ [0.5ex]
   \hline
   \rule{0pt}{2.5ex}$\chi^2_\mathrm{tot}$ & $1987$ & $1988.3$ & $1978.8$ & $1978.9$ & $1982$ & $1982.6$ & $1978.6$ & $1979$ \\ [0.5ex]
   \hline
   \hline
   \rule{0pt}{2.5ex}planck\_2018\_lowl.EE\_sroll2 & $-$ & $-$ & $-$ & $-$ & $-$ & $-$ & $-$ & $-$ \\[0.5ex]
   act\_dr6\_cmbonly & $155.7$ & $156.1$ & $154.6$ & $155.5$ & $155.5$ & $156.5$ & $154.4$ & $155.3$ \\[0.5ex]
   bao.desi\_dr2.desi\_bao\_all & $11.1$ & $11.1$ & $8.9$ & $8.7$ & $10$ & $9.5$ & $9.2$ & $8.8$ \\ [0.5ex]
   sn.pantheonplus & $1406.3$ & $1406.2$ & $1403.1$ & $1402.7$ & $1402.8$ & $1402.6$ & $1402.8$ & $1402.8$ \\ [0.5ex]
   act\_dr6\_lenslike.ACTDR6LensLike & $19.4$ & $19.5$ & $19.4$ & $19.9$ & $19.3$ & $19.5$ & $19.4$ & $19.7$ \\ [0.5ex]
   \hline
   \rule{0pt}{2.5ex}$\chi^2_\mathrm{tot}$ & $1592.4$ & $1592.9$ & $1585.9$ & $1586.8$ & $1587.6$ & $1588.2$ & $1585.8$ & $1586.6$ \\ [0.5ex]
   \hline
   \hline
   \end{tabular}
}
\end{table}

\begin{table}[htbp!]
\caption{$\chi^2_\mathrm{min}$ contribution from each dataset for Planck CMB data. The first column for each parameterization is the bestfit $\chi^2$ values while the second column is the MAP $\chi^2$ values. Upper part of the table is for dataset combination including low-$\ell$ EE and lower part is excluding low-$\ell$ EE.}
\label{table5}
\centering
\resizebox{\textwidth}{!}{
   \begin{tabular}{l|c c|c c|c c|c c}
   \hline
   \thead{Dataset} & \multicolumn{2}{c|}{\thead{$\Lambda$CDM}} & \multicolumn{2}{c|}{\thead{CPL}} & \multicolumn{2}{c|}{\thead{JBP}} & \multicolumn{2}{c}{\thead{BA}} \\
   \hline
   \hline
   \rule{0pt}{2.5ex}planck\_2018\_lowl.EE\_sroll2 & $390.1$ & $391.3$ & $389.9$ & $390.2$ & $390.3$ & $390.4$ & $390.1$ & $390.2$ \\[0.5ex]
   planck\_2018\_lowl.TT & $22.7$ & $23$ & $23.2$ & $23$ & $22.8$ & $23.1$ & $23.1$ & $23.2$ \\[0.5ex]
   planck\_NPIPE\_highl\_CamSpec.TTTEEE & $10545.3$ & $10544.4$ & $10542.5$ & $10543$ & $10544.1$ & $10543.1$ & $10542.5$ & $10542.2$ \\[0.5ex]
   bao.desi\_dr2.desi\_bao\_all & $12.3$ & $13.7$ & $9.6$ & $10.1$ & $10.8$ & $12$ & $9.6$ & $9.5$ \\ [0.5ex]
   sn.pantheonplus & $1405.6$ & $1405.3$ & $1403.3$ & $1403.2$ & $1403.1$ & $1403.4$ & $1402.9$ & $1403.3$ \\ [0.5ex]
   act\_dr6\_lenslike.ACTDR6LensLike & $19.8$ & $20.6$ & $19.6$ & $19.9$ & $19.7$ & $19.9$ & $19.7$ & $19.7$ \\ [0.5ex]
   \hline
   \rule{0pt}{2.5ex}$\chi^2_\mathrm{tot}$ & $12395.8$ & $12398.3$ & $12388.1$ & $12389.4$ & $12390.7$ & $12391.8$ & $12387.8$ & $12388.1$ \\ [0.5ex]
   \hline
   \hline
   \rule{0pt}{2.5ex}planck\_2018\_lowl.EE\_sroll2 & $-$ & $-$ & $-$ & $-$ & $-$ & $-$ & $-$ & $-$ \\[0.5ex]
   planck\_2018\_lowl.TT & $23.3$ & $23.2$ & $23.3$ & $23.2$ & $23.4$ & $23.5$ & $23.2$ & $23.4$ \\[0.5ex]
   planck\_NPIPE\_highl\_CamSpec.TTTEEE & $10544.1$ & $10544.5$ & $10542.2$ & $10542.8$ & $10543.8$ & $10543.3$ & $10542.5$ & $10542.1$ \\[0.5ex]
   bao.desi\_dr2.desi\_bao\_all & $11.2$ & $10.9$ & $9.1$ & $9.4$ & $9.7$ & $10.3$ & $9.1$ & $9$ \\ [0.5ex]
   sn.pantheonplus & $1406$ & $1406.2$ & $1403.1$ & $1402.7$ & $1403$ & $1402.8$ & $1403$ & $1403.2$ \\ [0.5ex]
   act\_dr6\_lenslike.ACTDR6LensLike & $19.5$ & $19.5$ & $19.6$ & $19.5$ & $19.5$ & $19.5$ & $19.5$ & $19.6$ \\ [0.5ex]
   \hline
   \rule{0pt}{2.5ex}$\chi^2_\mathrm{tot}$ & $12004.1$ & $12004.3$ & $11997.3$ & $11997.6$ & $11999.3$ & $11999.4$ & $11997.3$ & $11997.2$ \\ [0.5ex]
   \hline
   \hline
   \end{tabular}
}
\end{table}

\section{Results using DESY5 SNe dataset}
\label{appC}

\begin{figure}[H]
    \centering
    \subfloat[CPL\label{10a}]{\includegraphics[width=0.33\textwidth,keepaspectratio]{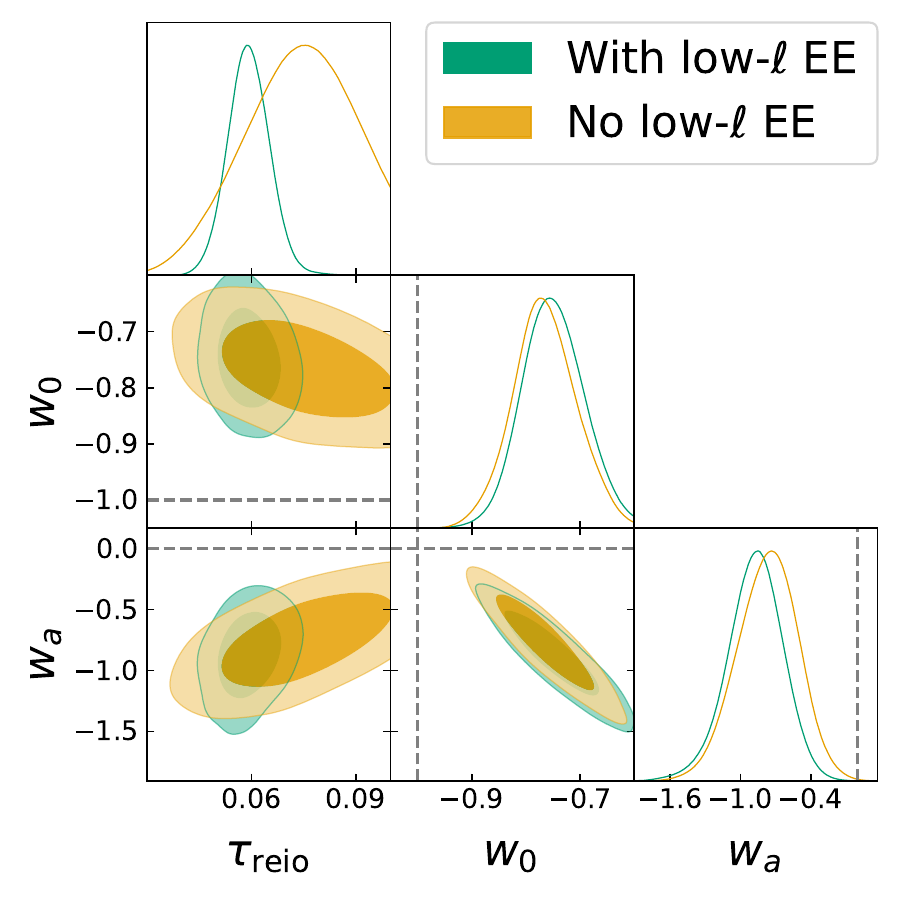}}
    \subfloat[JBP\label{10b}]{\includegraphics[width=0.33\textwidth,keepaspectratio]{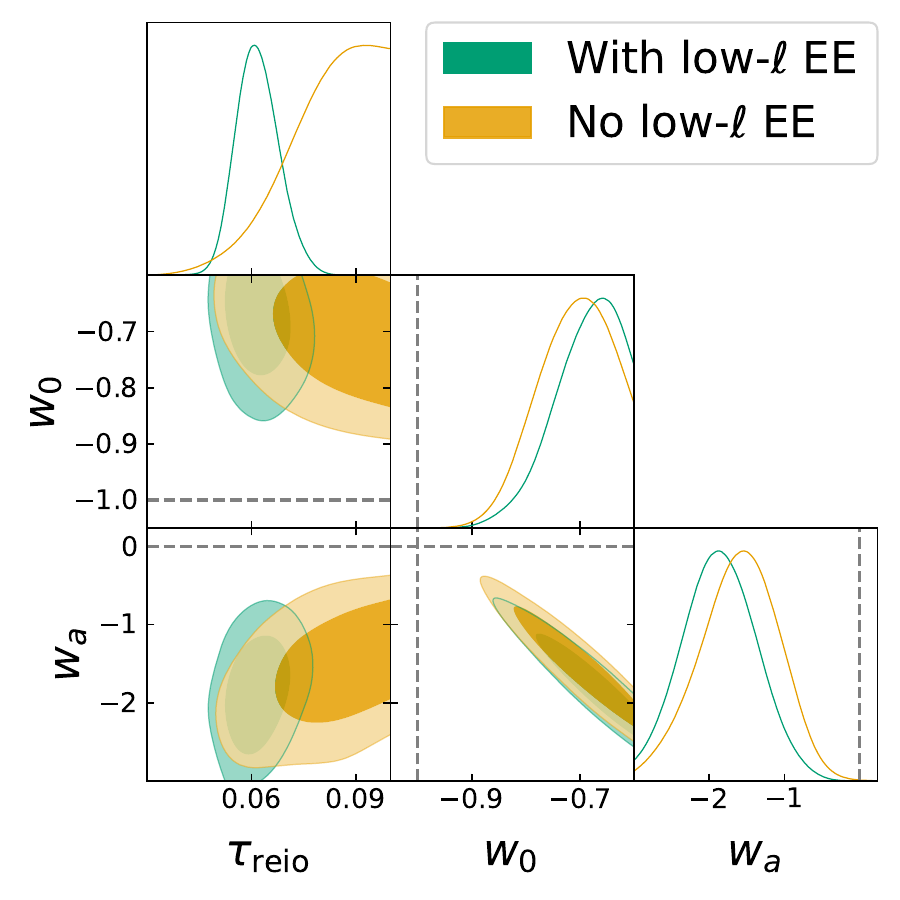}}
    \subfloat[BA\label{10c}]{\includegraphics[width=0.33\textwidth,keepaspectratio]{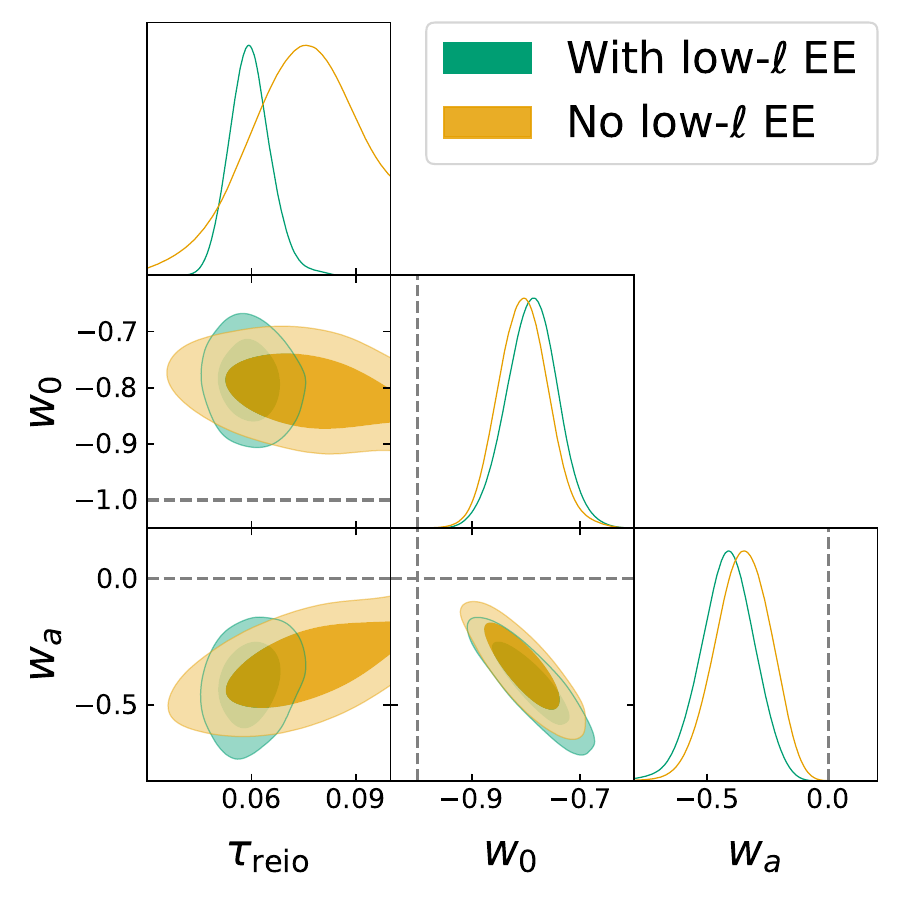}} \\
    \subfloat[EOS for CPL\label{10d}]{\includegraphics[width=0.33\textwidth,keepaspectratio]{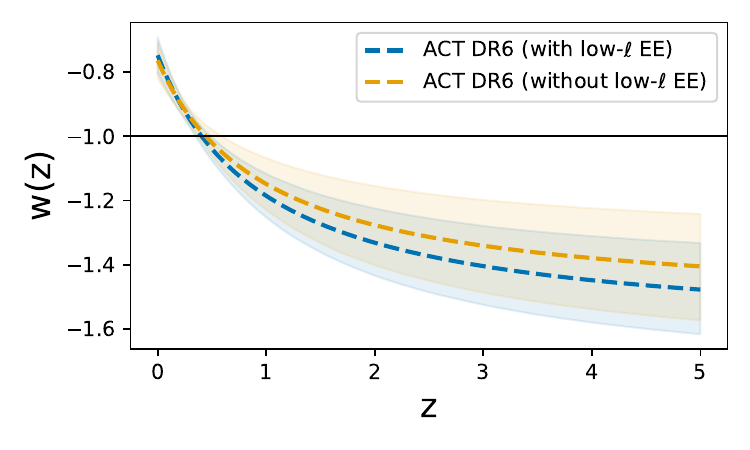}}
    \subfloat[EOS for JBP\label{10e}]{\includegraphics[width=0.33\textwidth,keepaspectratio]{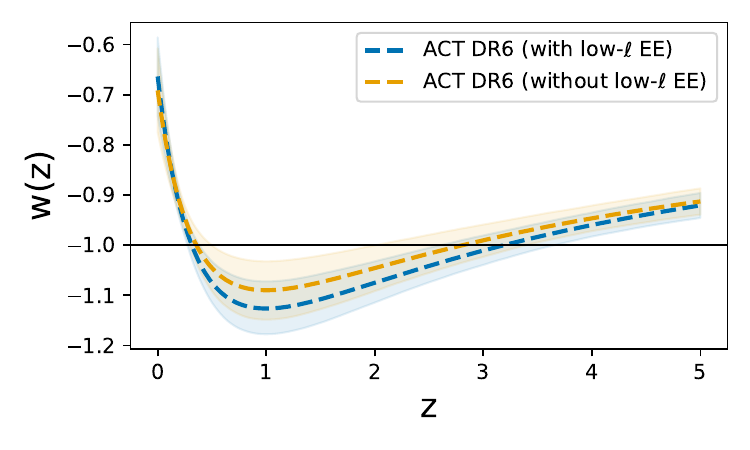}}
    \subfloat[EOS for BA\label{10f}]{\includegraphics[width=0.33\textwidth,keepaspectratio]{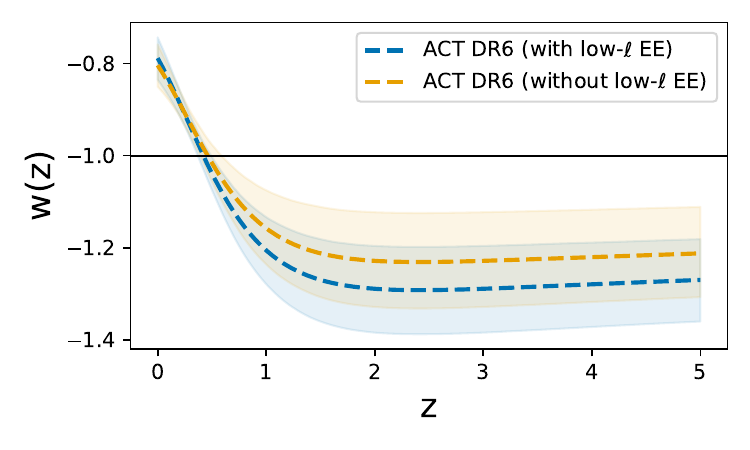}} \\
    \caption{Figs.~\ref{10a}, \ref{10b} and \ref{10c} represent the $68\%$ and $95\%$ credible intervals for ACT DR6$+$BAO$+$DESY5. Figs.~\ref{10d}, \ref{10e} and \ref{10f} show the EOS for the corresponding parameterizations.}
    \label{fig10}
\end{figure}

We present the results when using the ACT DR6$+$BAO$+$DESY5 SNe dataset combination in Table~\ref{table6} and Fig.~\ref{fig10}. We recover the main results found when using the PP SNe compilation. $n_s$ and $\Omega_bh^2$ are insensitive to the inclusion of low-$\ell$ EE data. When excluding low-$\ell$ EE data, there is a mild upward shift in $A_s$, which is larger in $\Lambda$CDM ($1.6\sigma$) and JBP ($1.39\sigma$) and smaller in CPL ($0.8\sigma$) and BA ($0.9\sigma$). In $\Lambda$CDM the $\sigma_8$ values shift upward ($1.55\sigma$) when excluding low-$\ell$ EE data due to large error bars (relaxation of $\sim43\%$). However, for CPL, JBP and BA, the discrepancy in $\sigma_8$ values is  within $0.6\sigma$ when excluding low-$\ell$ EE data.

Contrary to including the PP SNe compilation (Figs.~\ref{2b}, \ref{3b} and \ref{4b}), the joint $(w_0, w_a)$ constraints show a large discrepancy ($>2 \sigma$) with the standard $\Lambda$CDM value of $(-1, 0)$. The EOS shows a similar trend of moving close to the quintessence regime when excluding the large-scale EE polarization data.

\begin{table}[htbp!]
\caption{$68\%$ credible intervals for cosmological parameters for ACT DR6 CMB$+$BAO$+$DESY5 dataset combination.}
\label{table6}
\centering
\resizebox{\textwidth}{!}{
    \begin{tabular}{l|c|c|c|c|c|c|c|c}
    \hline
    \thead{Parameters} & \multicolumn{2}{c|}{\thead{$\Lambda$CDM}} & \multicolumn{2}{c|}{\thead{CPL}} & \multicolumn{2}{c|}{\thead{JBP}} & \multicolumn{2}{c}{\thead{BA}} \\
    & \thead{low-$l$ EE} & \thead{no low-$l$ EE} & \thead{low-$l$ EE} & \thead{no low-$l$ EE} & \thead{low-$l$ EE} & \thead{no low-$l$ EE} & \thead{low-$l$ EE} & \thead{no low-$l$ EE} \\
    \hline
    \hline
    \rule{0pt}{3ex}$100\Omega_bh^2$ & $2.261\pm0.016$ & $2.259\pm0.016$ & $2.259\pm0.016$ & $2.260\pm0.016$ & $2.261\pm0.016$ & $2.261\pm0.016$ & $2.259\pm0.016$ & $2.259\pm0.016$ \\[1.5ex]
    $100\Omega_ch^2$ & $11.812\pm0.072$ & $11.766\pm0.069$ & $11.96\pm0.10$ & $11.88\pm0.13$ & $11.88^{+0.11}_{-0.095}$ &  $11.79^{+0.11}_{-0.12}$ & $11.96\pm0.10$ & $11.88\pm0.13$ \\[1.5ex]
    $n_s$ & $0.9750\pm0.0063$ & $0.9786\pm0.0063$ & $0.9739^{+0.0058}_{-0.0065}$ & $0.9766^{+0.0060}_{-0.0068}$ & $0.9741\pm0.0063$ & $0.9779\pm0.0065$ & $0.9763\pm0.0063$ & $0.9767\pm0.0069$ \\[1.5ex]
    $A_s (10^{-9})$ & $2.132^{+0.022}_{-0.025}$ & $2.205^{+0.045}_{-0.029}$ & $2.111^{+0.022}_{-0.025}$ & $2.159^{+0.060}_{-0.051}$ & $2.124^{+0.023}_{-0.026}$ & $2.195^{+0.055}_{-0.036}$ & $2.113\pm0.023$ & $2.163^{+0.059}_{-0.048}$ \\[1.5ex]
    $\tau_\mathrm{reio}$ & $0.0629^{+0.0052}_{-0.0066}$ & $0.0841^{+0.015}_{-0.0044}$ & $0.0592\pm0.0060$ & $0.073^{+0.018}_{-0.013}$ & $0.0618^{+0.0055}_{-0.0066}$ & $>0.0773$ & $0.0596^{+0.0053}_{-0.0062}$ & $0.074^{+0.017}_{-0.013}$ \\[1.5ex]
    $z_\mathrm{reio}$ & $8.44^{+0.52}_{-0.62}$ & $10.4^{+1.1}_{-0.54}$ & $8.09\pm0.59$ & $9.3^{+1.6}_{-1.1}$ & $8.33^{+0.55}_{-0.62}$ & $10.1^{+1.4}_{-0.52}$ & $8.13\pm0.58$ & $9.4^{+1.5}_{-1.1}$ \\[1.5ex]
    $H_0$[km/s/Mpc]  & $68.26\pm0.29$ & $68.41\pm0.28$ & $66.93\pm0.59$ & $66.91\pm0.54$ & $66.79\pm0.55$ & $66.74\pm0.58$ & $67.00\pm0.55$ & $66.95\pm0.58$ \\[1.5ex]
    $\sigma_8$ & $0.8130\pm0.0049$ & $0.8263^{+0.0084}_{-0.0057}$ & $0.8113\pm0.0089$ & $0.8153\pm0.0094$ & $0.8054\pm0.0086$ & $0.8123\pm0.0094$ & $0.8118\pm0.0088$ & $0.8158\pm0.0097$ \\[1.5ex]
    $w_0$ & $-$ & $-$ & $-0.748^{+0.054}_{-0.061}$ & $-0.765\pm0.059$ & $-0.664\pm0.078$ & $-0.691\pm0.082$ & $-0.788\pm0.047$ & $-0.804\pm0.047$ \\[1.5ex]
    $w_a$ & $-$ & $-$ & $-0.88^{+0.24}_{-0.22}$ & $-0.77^{+0.27}_{-0.24}$ & $-1.85\pm0.48$ & $-1.60^{+0.56}_{-0.49}$ & $-0.35^{+0.12}_{-0.11}$ & $-0.42\pm0.11$ \\[1.5ex]
    \hline
    \end{tabular}
    }
\end{table}

\bibliography{references}

\end{document}